\documentclass[AMA,LATO1COL]{WileyNJD-v2}
\usepackage{moreverb}

\newcommand\BibTeX{{\rmfamily B\kern-.05em \textsc{i\kern-.025em b}\kern-.08em
T\kern-.1667em\lower.7ex\hbox{E}\kern-.125emX}}

\articletype{Article Type}%

\received{<day> <Month>, <year>}
\revised{<day> <Month>, <year>}
\accepted{<day> <Month>, <year>}

\begin{document}

\title{Are your comments outdated? A method for code-comment consistency detection}
\title{Are your comments outdated? Towards automatically detecting code-comment consistency}

\author[1]{Yuan Huang}

\author[1]{Yinan Chen}

\author[2]{Xiangping Chen*}

\author[3]{Xiaocong Zhou}

\authormark{Yuan Huang \textsc{et al}}

\address[1]{\orgdiv{School of Software Engineering}, \orgname{Sun Yat-sen University}, \orgaddress{\state{Guangdong}, \country{China}}}

\address[2]{\orgdiv{School of Communication and Design}, \orgname{Sun Yat-sen University}, \orgaddress{\state{Guangdong}, \country{China}}}

\address[3]{\orgdiv{School of Computer Science and Engineering}, \orgname{Sun Yat-sen University}, \orgaddress{\state{Guangdong}, \country{China}}}

\corres{*Corresponding author name, Corresponding address. \email{chenxp8@mail.sysu.edu.cn}}

\presentaddress{Sun Yat-sen University, Guangzhou, 510000, Guangdong, China}

\abstract[Abstract]{In software development and maintenance, code comments can help developers understand source code, and improve communication among developers. However, developers sometimes neglect to update the corresponding comment when changing the code, resulting in outdated comments (i.e., inconsistent codes and comments). Outdated comments are dangerous and harmful and may mislead subsequent developers. More seriously, the outdated comments may lead to a fatal flaw sometime in the future. To automatically identify the outdated comments in source code, we proposed a learning-based method, called CoCC, to detect the consistency between code and comment.  To efficiently identify outdated comments, we extract multiple features from both codes and comments before and after they change. Besides, we also consider the relation between code and comment in our model. Experiment results show that CoCC can effectively detect outdated comments with precision over 90\%. In addition, we have identified the 15 most important factors that cause outdated comments, and verified the applicability of CoCC in different programming languages. We also used CoCC to find outdated comments in the latest commits of open source projects, which further proves the effectiveness of the proposed method.} 


% \jnlcitation{\cname{%
% \author{Williams K.}, 
% \author{B. Hoskins}, 
% \author{R. Lee}, 
% \author{G. Masato}, and 
% \author{T. Woollings}} (\cyear{2016}), 
% \ctitle{A regime analysis of Atlantic winter jet variability applied to evaluate HadGEM3-GC2}, \cjournal{Q.J.R. Meteorol. Soc.}, \cvol{2017;00:1--6}.}

\maketitle

%\footnotetext{\textbf{Abbreviations:} ANA, anti-nuclear antibodies; APC, antigen-presenting cells; IRF, interferon regulatory factor}

\section{Introduction}\label{sec1}

Guaranteeing the software quality is critical, because a minor error may lead to a serious consequence \cite{ibrahim2012relationship,huang2020towards}. In general, most of the previous studies focus on the code quality of software, such as code reliability \cite{parnas2011precise,keyes2002software}, code vulnerability \cite{alarcon2020would,yang2021source,semasaba2022empirical}, etc. While the qualities of other software artifacts are also important, such as the quality of code documents, e.g, code comments \cite{rani2021speculative,pascarella2019classifying,gosling2000Java,fluri2007code}. 

Code comments record various code information, such as why and how functions are implemented, how APIs are used, how code segments relate to each other, and so on \cite{kuang2022code,shahbazi2021api2com}. 
Code comments improve the readability of the codes and express the programmer's intent in a clearer way \cite{kuang2022code,zhu2022simple}, which further helps programmers understand the source code and improve communication between developers \cite{shahbazi2021api2com,mastropaolo2021empirical}. Code comments play an important role in software development and maintenance \cite{zhu2022simple,steidl2013quality,fluri2009analyzing,hu2020deep}.

However, the quality of comments may not be  guaranteed at all times. In software development, programmers often change the code to fix a bug or add new functionality, while the comments corresponding to the codes may not be updated in time. 

Comments that are not updated can cause inconsistencies with the code, and we call them outdated comments. Outdated comments  lose their timeliness, which may mislead subsequent developers \cite{ibrahim2012relationship,parnas2011precise}. A previous study also shows that outdated comments can lead to a defect in the software system \cite{ibrahim2012relationship}.  

\begin{figure}[htbp]
\centering
\includegraphics[scale=0.4]{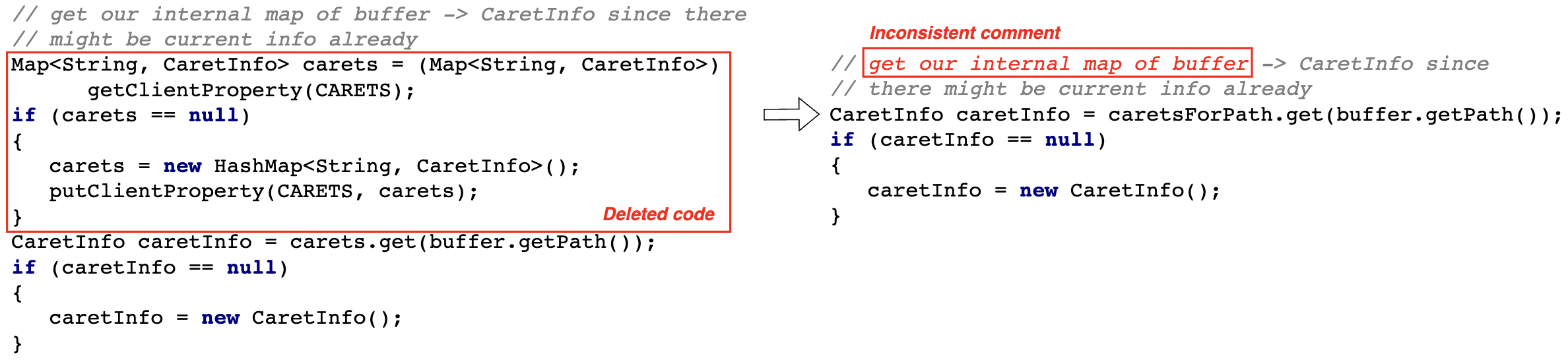}
\caption{Comment contains a description of the code that was removed in jEdit commit \#13416.}
\label{out1}
\end{figure}

\begin{figure}[htbp]
\centering
\includegraphics[scale=0.45]{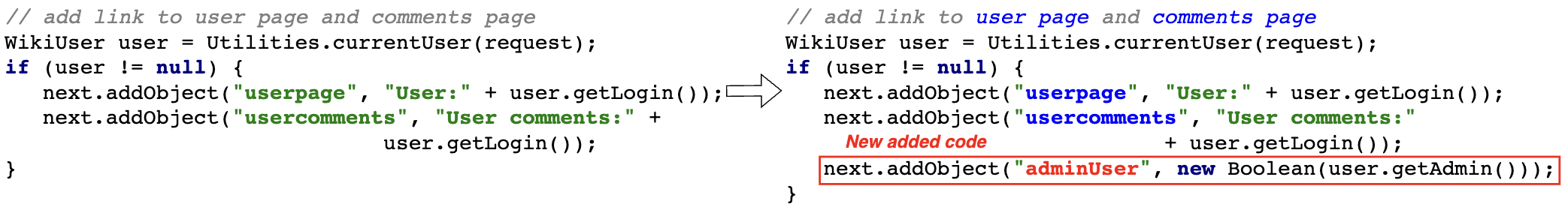}
\caption{Comment that lacks a description of the new code in JAMWiki commit \#304.}
\label{out2}
\end{figure}

\begin{figure}[htbp]
\centering
\includegraphics[scale=0.4]{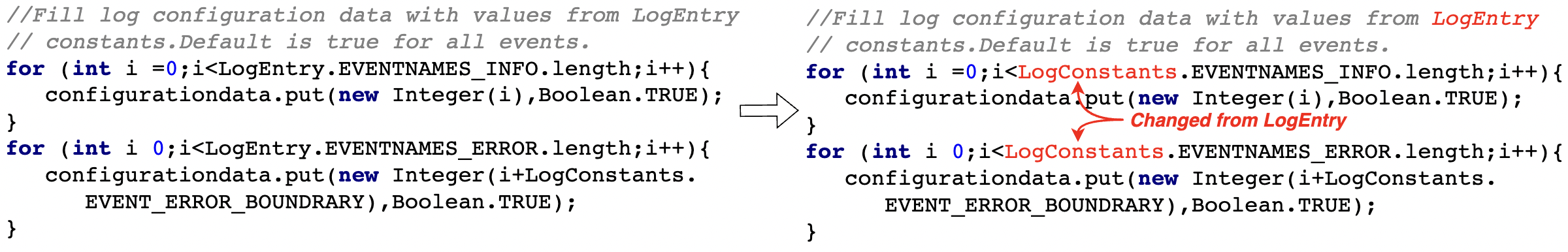}
\caption{Comment that refers to variables that do not exist in EJBCA commit \#4977.}
\label{out3}
\end{figure}

Most outdated comments are due to the code changes \cite{fluri2007code}. FIGURE \ref{out1} to FIGURE \ref{out3} shows examples of outdated comments during the code change. The comment in FIGURE \ref{out1} contains a description of the removed code. The comment in FIGURE \ref{out2} lacks a description of the new code, which may mislead the code reader about the function implementation. The comment in FIGURE \ref{out3} refers to variables that do not exist, which may confuse the code reader.

\begin{figure}[htbp]
\centering
\includegraphics[scale=0.3]{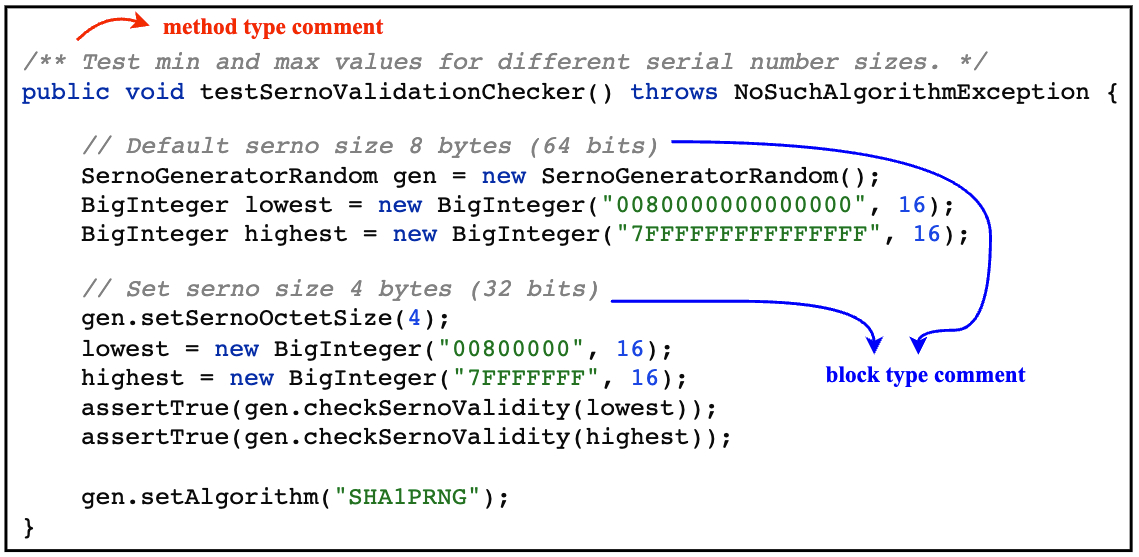}
\caption{Method-type comment and block-type comment.}
\label{commenttype}
\end{figure}

In this paper, we focus on two types of comments: method-type comments and block-type comments \cite{huang2020towards,fluri2007code,huang2020does}. Method-type comments are comments used in method headers to describe the functionality of methods  (also called header comments, Javadoc). Block-type comments are comments within the method or class body that describe the code lines inside the method or class.  Method-type comments have a clear commenting scope, i.e., the whole method, while block-type comments may have a commenting scope of one or several code lines \cite{Huanchao2019}, and we will use heuristic rules to define the scope of block-type comments (detailed in Section 2.1). FIGURE \ref{commenttype} shows these two types of comments.

\begin{figure}[htbp]
\centering
\includegraphics[scale=0.3]{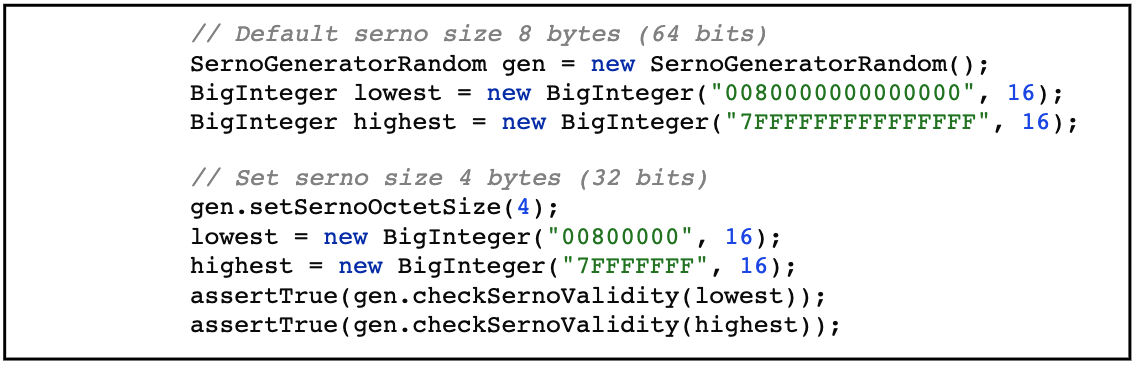}
\caption{Two block-type code-comment pair extracted from FIGURE \ref{commenttype}.}
\label{blockc}
\end{figure}

\begin{figure}[htbp]
\centering
\includegraphics[scale=0.3]{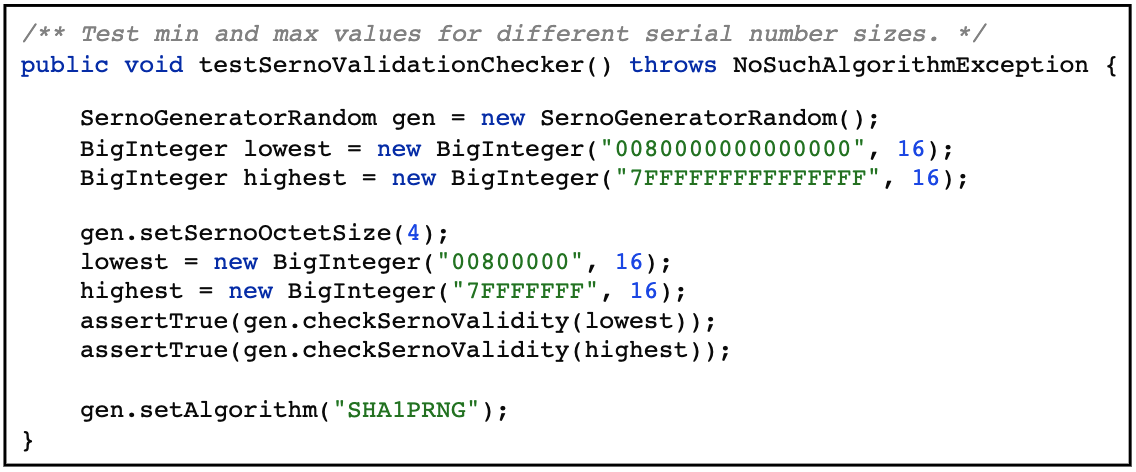}
\caption{Method-type code-comment pair extracted from FIGURE \ref{commenttype}.}
\label{methodb}
\end{figure}

The quality analysis of code comments by previous studies mainly focuses on the relation between code and comment \cite{ibrahim2012relationship,fluri2007code,fluri2009analyzing}, and they point out the importance of maintaining consistency between code and comment. In this study, to efficiently identify the outdated comments, we propose CoCC to detect the \textbf{co}nsistency between \textbf{c}ode and \textbf{c}omment. We first collect a large-scale dataset from 22 open-source Java projects and identify the outdated and un-outdated comments from the commits of these projects. Then, we extract multiple code features, comment features, and relation features between comment and code. We trained several machine learning models to identify outdated comments. Experiment results show that CoCC can effectively detect outdated comments with precision over 90\%. In addition, we used CoCC to find out the outdated comments in the latest commits of the open-source projects. These outdated comments are really outdated after manual checks by programmers. To verify the applicability of CoCC in different languages, we extended the experiment to Python, and CoCC also performed well.

To facilitate research and application, our replication package\footnote{\url{https://github.com/chenyn273/CoCC}} and dataset\footnote{\url{https://drive.google.com/drive/folders/12xYfd8DC66OdBy3HhZs2T8qrGBfVEPCo?usp=sharing}} are released.

The main contributions of this study are as follows:
\begin{itemize}
\item CoCC is designed to identify the outdated comments from source code, which is suitable for method-type comments and block-type comments.
\item A dataset for outdated comment detection is published.
\item We propose useful features (i.e., code features, comment features, and relation features) to detect whether comments are outdated.
\item The comprehensive evaluation results demonstrate the feasibility and effectiveness of CoCC. The result of the extended experiment on Python proved its applicability in different programming languages.

\end{itemize}

A previous version\cite{liu2018automatic} of this work was published at the 42nd IEEE COMPSAC, and this paper significantly extends it. The previous version focused only on detecting outdated block-type comments. In this paper, we detect both outdated block-type comments and method-type comments. In addition, we extended the dataset to 83,916 changes (35,050 changes in the previous version). Besides, we extended the features and removed the highly correlated features (detailed in Appendix A), and the performance of CoCC was significantly improved. In addition, we evaluated the performance of random forest\cite{ho1995random} and other machine learning models (i.e., XGBoost\cite{2016XGBoost}, logistic regression\cite{dreiseitl2002logistic}, naive bayes, SVM\cite{cortes1995support} and decision tree\cite{song2015decision}) in outdated comment detection task. We also extended the experiment on Python to verify the applicability of CoCC for different programming languages.

The rest of this paper is organized as follows. Section 2 details the proposed method. The experimental setup is discussed in Section 3. Section 4 is the result discussion.
Section 5 surveys and summarizes related research work. Section 6 is the threat to validity. Finally, Section 7 concludes the paper and points out possible future directions.

\section{APPROACH}

\begin{figure}[htbp]
\centering
\includegraphics[scale=0.4]{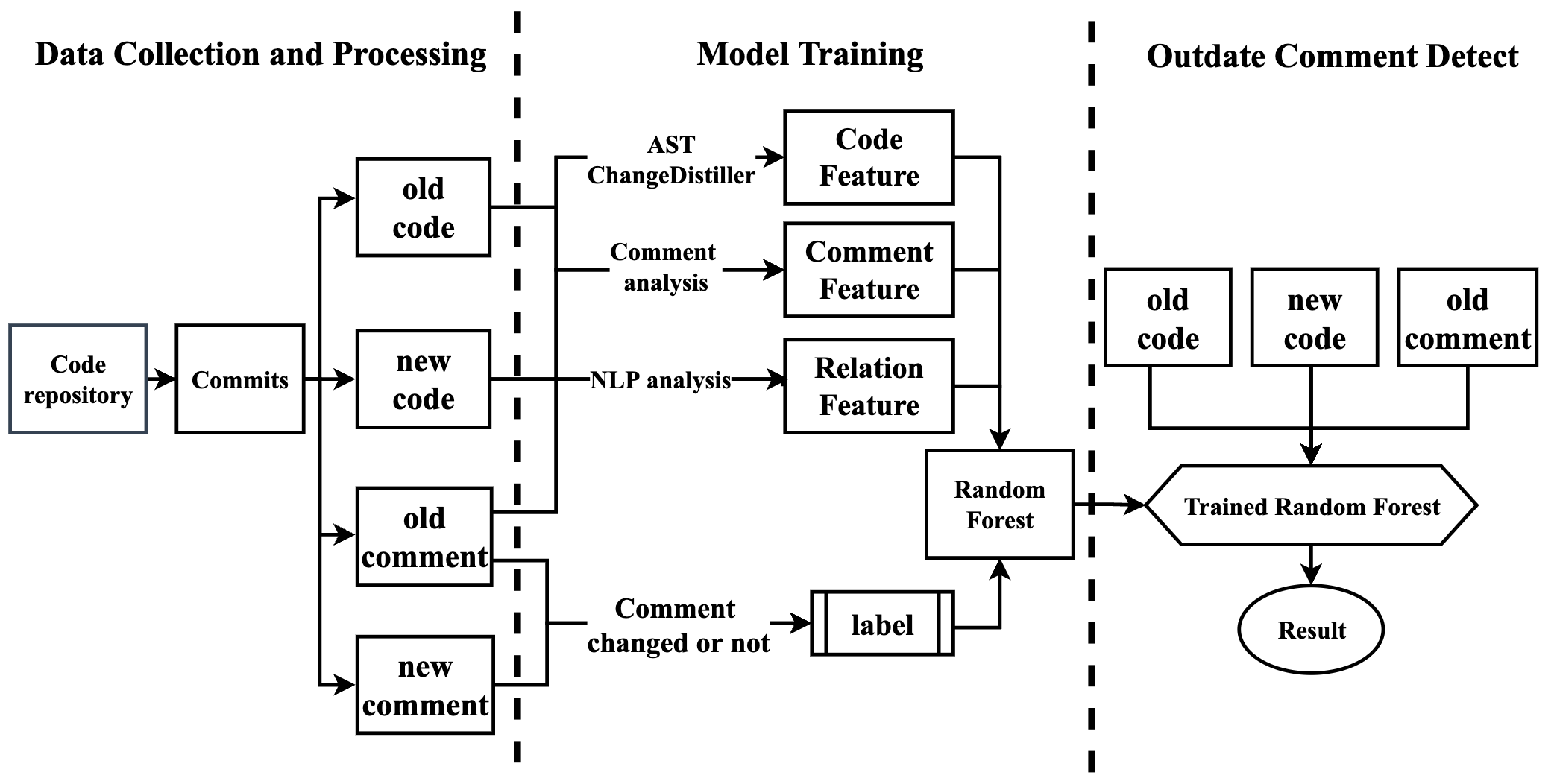}
\caption{Overall architecture of CoCC.}
\label{overall}
\end{figure}

The overall architecture of CoCC is shown in FIGURE \ref{overall}. It consists of three stages: data collection and processing, model training, and outdated comment detection. 
Specifically, in data collection, we collect the code-comment pairs from the commits of the open-source projects. We require that each commit contains at least one old code, new code, associated old comment, and associated new comment. Depending on whether the old comment and new comment are the same, we labeled each code-comment pair as a positive sample (old comment and new comment are different with label "1", outdated comment) and a negative sample (old comment and new comment are the same with a label "0", not outdated comment). Then, we extract code features, comment features, and relation features from the code-comment pairs using ChangeDistiller\cite{4339230}, NLP and AST tools, and use these features to train the classification model. Finally, given the old and new codes and the associated old comments, we can use the trained model to predict whether the old comments need to be updated or not.

%In this section, we first describe how to collect and process data, then how to select and extract features, and finally how to train the model.

\subsection{Code change extraction}\label{AA}

The goal of this study is to determine whether the comments need to be updated when the code is changed. Therefore, we need to extract the code-comment pairs, we focus on block-type (shown in FIGURE \ref{blockc}) and method-type (shown in FIGURE \ref{methodb}) comments in this study. In software development, the commits record the code change information of the projects, and we can extract the code-comment pairs from the commits of each project.

To collect block-type code-comment pairs, we use the following heuristic rules to link code and comment: (1) Adjacent block-type comments are treated as one comment since there is no code between them. (2) The scope of a comment starts with the first line of code after the comment. (3) The scope of a comment ends with the last line of code before another comment in the same block (not a sub-block), or at the end of a block or method. Two block-type code-comment pairs extracted from FIGURE \ref{commenttype} using heuristic rules are shown in FIGURE \ref{blockc}.

To collect method-type code-comment pairs, we use AST tools to get the comment and related method body and remove the irrelevant comments in the method body. For example, when extracting the method-type code-comment pair in FIGURE \ref{commenttype}, we use ChangeDistiller to get the comment and the method body, then we removed the irrelevant comments "\textbf{//Default serno size 8 bytes(64bits)}" and "\textbf{//Set serno size 4 bytes(32 bits)}" in the method body, then get the method-type code-comment pair shown as FIGURE \ref{methodb}. 

After collecting the code-comment pairs before and after the code change, we use the code change extraction tool ChangeDistiller to extract changes in the code-comment pairs. The tool can output fine-grained change information by taking the code-comment pair before the change and the code-comment pair after the change as the inputs.

In addition, the heuristic rules to link code and comment may inaccurate, which will affect the quality of the block-type code-comment pair dataset. We take this as a threat to validity (detailed in threat to validity).

\subsection{Feature extraction}

This section describes how to extract features from codes and comments. Whether a comment is outdated can be determined by the code change information, the comment information, and the relation between the code and the comment, we extract features from these three dimensions. Considering that there are too many features, to prevent overfitting and reduce the correlation of features, we calculated the correlation of features and removed the high correlation features (detailed in Appendix A). The features introduced below are retained after filtering. Most of these features can be extracted by writing programs or using the outputs of ChangeDistiller. We will introduce feature extraction methods in the last paragraph of each feature subsection.

\subsubsection{Code features}
Code features describe the code change information and the state of the code before and after the change. We extract features at the class, method, and statement levels. TABLE \ref{table:codefeature} shows the detailed code features.

% \usepackage{multirow}
\begin{table}[htbp]
\centering
\caption{Code features.}
\label{table:codefeature}
\begin{tabular}{|l|ll|l|}
\hline
\textbf{Level} &
  \multicolumn{2}{l|}{\textbf{Feature}} &
  \textbf{Description} \\ \hline
\multirow{2}{*}{Class} &
  \multicolumn{2}{l|}{\multirow{2}{*}{Class attributes change}} &
  \multirow{2}{*}{Whether the class attributes are modified?} \\
 &
  \multicolumn{2}{l|}{} &
   \\ \hline
\multirow{3}{*}{Method} &
  \multicolumn{1}{l|}{\multirow{3}{*}{Method declaration change}} &
  Method name &
  Whether the method name is modified? \\ \cline{3-4} 
 &
  \multicolumn{1}{l|}{} &
  Return type &
  Whether the return type is modified? \\ \cline{3-4} 
 &
  \multicolumn{1}{l|}{} &
  Parameter &
  Whether the parameters are modified? \\ \hline
\multirow{9}{*}{Statement} &
  \multicolumn{2}{l|}{Proportion of code lines in code-comment pair} &
  The ratio of code lines to code-comment pair lines. \\ \cline{2-4} 
 &
  \multicolumn{2}{l|}{Proportion of changed statement lines in code-comment pair} &
  \begin{tabular}[c]{@{}l@{}}The ratio of changed statement lines to\\  code-comment pair lines.\end{tabular} \\ \cline{2-4} 
 &
  \multicolumn{1}{l|}{\multirow{3}{*}{\begin{tabular}[c]{@{}l@{}}Statement changes (Statement type: IF, \\ ELSE IF, FOR, WHILE, CATCH, TRY, THROW, METHOD \\ INVOCATION,  VARIABLE DECLARATION)\end{tabular}}} &
  Add &
  Add a statement. \\ \cline{3-4} 
 &
  \multicolumn{1}{l|}{} &
  Delete &
  Delete a statement. \\ \cline{3-4} 
 &
  \multicolumn{1}{l|}{} &
  Update &
  Update a statement. \\ \cline{2-4} 
 &
  \multicolumn{2}{l|}{Refactoring} &
  Whether the code change includes code refactoring? \\ \cline{2-4} 
 &
  \multicolumn{2}{l|}{\multirow{3}{*}{Code word analysis}} &
  \multirow{3}{*}{\begin{tabular}[c]{@{}l@{}}The distance of the proportion of parts of speech in \\ the code before and after the change.\end{tabular}} \\
 &
  \multicolumn{2}{l|}{} &
   \\
 &
  \multicolumn{2}{l|}{} &
   \\ \hline
\end{tabular}

\end{table}

In programs written in object-oriented programming languages, the class and method declarations are in the context of the code change. In the class where the code change is located, if class attributes are changed and the changed class attributes are used in the changed code, the change in the attribute may affect the functionality implementation, and the comment should be changed accordingly. Similarly, in the method where the changed code is located, if the return type or parameters of the method are changed, the corresponding comments may be outdated.

The previous study \cite{liu2018automatic} has not investigated how different types of statement changes affect comment updates. Based on the commits collected from the projects (e.g., EJBCA, JAMWiki, JEdit, JHoDraw, OpenNMS, and so on), we explored different types of change statements that have different effects on comment updates. Adding and deleting WHILE statements, deleting ELSE-IF, enhanced FOR, and TRY statements will result in a more than 45$\%$ probability of outdated comments. In addition, adding FOR statements and deleting CATCH statements are more than 40$\%$ likely to cause outdated comments. This indicates that these listed changes are more relevant to outdated comments. Adding or deleting statements has a much greater impact on comment obsolescence than modifying statements. Therefore, we added statement changes features to the statement level features.

When code changes include refactoring, the code structure usually changes greatly, which may cause comments outdated. Refactoring is the process of changing a software system in such a way that it does not alter the external behavior of the code, yet improves the modular structure of the software \cite{mens2004survey}. Code refactoring generally involves a large range of code modifications. Nevertheless, the modified code would not change the behavior of the function. The associated comment may be kept unchanged if the comment describes the behavior without details. Therefore, in the feature selection, we selected eight types of code refactoring features, as listed in TABLE \ref{refactor}. In actual development, there are more than these eight types of refactoring, and we choose eight common and representative refactoring types according to the research about refactoring\cite{mens2004survey}.

\begin{table}[htb]
\centering
\caption{Refactoring features.}
\label{refactor}
\begin{tabular}{|l|l|}
\hline
\textbf{Refactoring} & \textbf{Description}                                                           \\ \hline
Extract method    & Turn the fragment into a method whose name explains the purpose of the method. \\ \hline
Inline method        & Put the method's body into the body of its callers and remove the method.      \\ \hline
Rename method      & Change the name of the method.                                                 \\ \hline
Add parameter        & Add a parameter for an object that can pass on the information.                \\ \hline
Remove parameter     & Remove the parameter which is no longer used by the method body.               \\ \hline
Inline temp &
  \begin{tabular}[c]{@{}l@{}}A temp that is assigned to once with a simple expression. Replace all references to that\\ temp with the expression.\end{tabular} \\ \hline
Encapsulate field    & There is a public field. Make it private and provide accessors.                \\ \hline
Introduce assertion &
  \begin{tabular}[c]{@{}l@{}}A section of code assumes something about the state of the program. Make the assumption \\ explicit with an assertion.\end{tabular} \\ \hline
\end{tabular}

\end{table}

\begin{algorithm}
\caption{"Extract method" detection method}\label{alg1}
\begin{algorithmic}
 \State extract changes using ChangeDistiller
 \State \textbf{for} each $Change$ \textbf{do}

\State \quad\; \textbf{if} $Change\_Type^{i}$ is $Statement\_Delete$ \textbf{and} $Change\_Type^{i+1}$ is $Method\_Invocation\_Insert$ \textbf{then}

\State \quad\;\quad\; $Old\_Code$ = use ChangeDistiller position information and AST to get old code
\State \quad\;\quad\; $New\_Code$ = match the method with the same method name and parameter list
\State \quad\;\quad\; $Old\_Code$ = $Old\_Code$ after removing the comments in it
\State \quad\;\quad\; $New\_Code$ = $New\_Code$ after removing the comments in it
\State \quad\; \textbf{end if}

\State \quad\; \textbf{if} $Old\_Code$ and $New\_Code$ matched \textbf{then}
\State  \quad\;\quad\; \textbf{return true}
\State \quad\; \textbf{end if}
\State \quad\; \textbf{if} $Old\_Code$ and $New\_Code$ didn't match \textbf{then}
\State \quad\;\quad\; \textbf{return false}
\State \quad\; \textbf{end if}
\State \textbf{end for}

\end{algorithmic}
\end{algorithm}

\begin{algorithm}
\caption{"Inline temp" detection method}\label{alg2}
\begin{algorithmic}
 \State extract changes using ChangeDistiller
 \State \textbf{for} each $Change$ \textbf{do}
    \State \quad\; \textbf{if} $Change\_Type^{i}$ is $Assignment\_Statement\_Delete$ \textbf{then}
        \State \quad\;\quad\; $Deleted\_Assigned\_Entity$ = use ChangeDistiller to get the deleted assigned entity
         \State \quad\;\quad\; $Deleted\_Assignment\_Entity$ = use ChangeDistiller to get the deleted assignment entity
      \State \quad\;\quad\;\textbf{for} $j>=i$ \textbf{do}
           \State \quad\;\quad\;\quad\; \textbf{if} $Change\_Type^{j}$ is $Statement\_Update$ \textbf{then}
                \State \quad\;\quad\;\quad\;\quad\; $Updated\_Entity$ = get updated entity with ChangeDistiller
                \State \quad\;\quad\;\quad\;\quad\; $Replaced\_Updated\_Entity$ = use ChangeDistiller to get the replaced updated entity
                \State \quad\;\quad\;\quad\; \textbf{end if}
           \State \quad\;\quad\;\quad\; \textbf{if} $Deleted\_Assigned\_Entity == Updated\_Entity$ \textbf{and} $Deleted\_Assignment\_Entity ==Replaced\_Updated\_Entity$ \textbf{then}
                \State \quad\;\quad\;\quad\;\quad\; \textbf{return true}
                \State \quad\;\quad\;\quad\; \textbf{end if}
                
            \State \quad\;\quad\;\quad\; $j=j+1$
\State \quad\;\quad\; \textbf{end for}
\State \quad\; \textbf{end if}
\State \textbf{end for}
\State \textbf{return false}
\end{algorithmic}
\end{algorithm}

Compared with the previous version, we considered the part of speech information of the code tokens. In general, changes in the noun or verb code tokens are more likely to make comments outdated. For example, in FIGURE \ref{blockc}, if the verb token "\textbf{set}" or noun token "\textbf{serno}" and "\textbf{size}" in "\textbf{setServoOctetSize(4)}" changed, the comment "\textbf{Set serno size 4 bytes (32bits)}" will be outdated. Therefore, we add the proportion of ten parts of speech in code before and after the change to the code features. We will discuss the effectiveness of word analysis features in Appendix B as well.

Code feature extraction method: (1) Use the number of code lines divided by the number of code-comment pair lines to get the "proportion of code lines in code-comment pair" feature. (2) Use the number of changed statement lines divided by the number of code-comment pair lines to get the "proportion of changed statement lines in code-comment pair" feature. (3) Use the Python tool, NLTK to get the "code word analysis" features, NLTK can do part of speech analysis. (4) Use the outputs of ChangeDistiller to get the rest features.

Refactoring feature extraction method: (1) See Algorithm \ref{alg1} to get the "extract method" refactoring feature. (2) "Inline method" is the inverse of "extract method", so the detection method is the inverse of "extract method" as well, see Algorithm \ref{alg1}. (3) See Algorithm \ref{alg2} to get the "inline temp" refactoring feature. (4) Use the outputs of ChangeDistiller to get the rest refactoring features.

\subsubsection{Comment features.}

\begin{table}[htb]
\centering
\caption{Comment features.}
\label{commentfeat}
\begin{tabular}{|l|l|}
\hline
\textbf{Feature}         & \textbf{Description}                                            \\ \hline
Todo comment    & Does the comment contain "TODO" information?           \\ \hline
Fix comment     & Does the comment contain "FIXME", or "FIXED" information? \\ \hline
Version comment & Does the comment contain "Version" information?        \\ \hline
Bug comment     & Does the comment contain "BUG" information?            \\ \hline
Word analysis &
  \begin{tabular}[c]{@{}l@{}}The proportion of parts of speech in the comment.\end{tabular} \\ \hline
\end{tabular}

\end{table}

TABLE \ref{commentfeat} lists the comment features. Special keywords are used in comments to indicate that codes and comments need to be updated in the next revision, such as task comments ("TODO", "FIXME", "XXX ") and error tags ("ERROR"). When updating code using such comments, the programmer should update the corresponding comments. For example, when a new function implementation is added in the expected location, the word "TODO" should be removed from the original comment to avoid confusion\cite{storey2008todo}. Similarly, the version type comments used to record the code version should be updated when the code change.

We also extended the comment features compared to the previous version. Similar to the code word analysis, the part of speech distribution in the comment also affects whether the comment is outdated. Comments are usually used to describe the behavior of code, which involves the corresponding number of verb tokens or noun tokens appearing in the comments. The greater the proportion of verb or noun tokens in comments, the greater the possibility that comments need to be updated in time when the code change. Therefore, we put the proportion of parts of speech in the comment into the comment features. We will discuss the effectiveness of word analysis features in Appendix B as well.

Comment feature extraction method: (1) Convert the comment to lowercase to determine whether it contains "todo", "fixme", "fixed", "bug", and "version". (2) Use the Python tool, NLTK to get the "word analysis" features, NLTK can do part of speech analysis.

\subsubsection{Relation features}
There are different guidelines on how to write useful comments\cite{steidl2013quality,mcburney2016empirical}. Comments should contain the intent and goal of the implementation\cite{steidl2013quality}, as well as possible additional insights behind the implementation. Thus, the description may refer to objects or features in the code fragment. In addition, a good comment written by developers should have a high semantic similarity to the source code.

In general, comments have a strong semantic association with the code snippet\cite{mcburney2016empirical}. If the similarity between the code and its corresponding comment changes significantly after a code change, the comment is likely to be updated. However, computing the similarity between code and comment is difficult because of the "lexical gap" between programming languages and natural languages. In this case, it is not possible to accurately measure similarity by counting common words contained in the code and comment or by directly embedding words. To address this problem, we employ a skip-gram model based on the approach proposed by Xin et al.\cite{ye2016word}, which bridges the lexical gap by projecting natural language utterances and code fragments as meaning vectors in shared representation space.

\begin{figure}[htbp]
\centering
\includegraphics[scale=0.33]{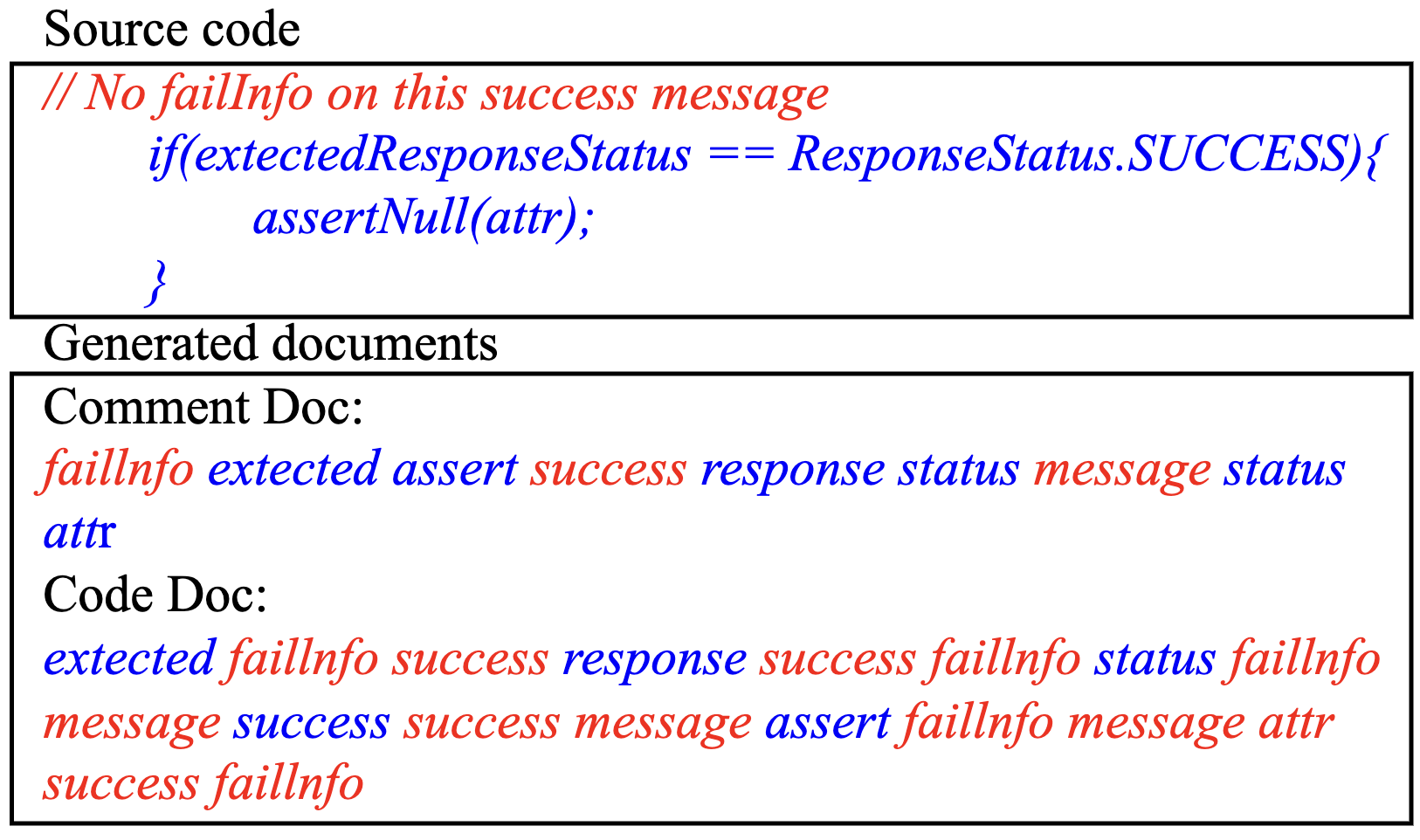}
\caption{Example of document generation}
\label{documentexmp}
\end{figure}

First, we preprocessed the source code and comment, including word splitting, stop word deletion, and stemming analysis. Second, for each word in the comment, we randomly selected two words from the code fragment and added them to the comment as a comment document. Third, for the code fragment, we randomly select two words from the associated comment and add them to the code document. Finally, these two documents are merged and serve as the corpus for our skip-gram model. FIGURE \ref{documentexmp} shows an example where, using the document generation rules, we generated two documents from the code fragment and its comment. The red words are generated from the comment and the blue words are from the code.

To obtain a vector representation of each word, we then trained the skip-gram model on the basis of the corpus (shown in FIGURE \ref{documentexmp}). Successive skip-grams are effective in predicting surrounding words in a contextual window of $2k+1$ words. The objective function of the skip-gram model aims to maximize the sum of the log probabilities of surrounding context words conditional on the central word\cite{mikolov2013distributed}.

\begin{equation}
\sum^{n}_{i=1} \sum^{}_{-kjk} \log p\left( w_{i+j}|w_{i}\right)  \label{eq}
\end{equation}

\noindent where $w_{i}$ and $w_{i+j}$ represent the central word and the surrounding context words in a context window of length $2k+1$, respectively, and $n$ represents the length of the word sequence. The term $\log p\left( w_{i+j}|w_{i}\right)$ is the conditional probability, which is defined using the softmax function.

\begin{equation}
\log p\left( w_{i+j}|w_{i}\right)  =\frac{\exp (v^{{}\prime^{T} }_{w_{i+j}}v_{w_{i}})}{\sum_{w\in W} \exp (v^{\prime^{T} \  }_{w}v_{w_{i}})}     
\end{equation}

\noindent where $v_{w}$ represents the input vector and $v^{\prime }_{w}$ represents the output vector of $w$ in the underlying neural model. $W$ represents the vocabulary of all words. Intuitively, $p(w_{i+j}|w_{i})$ estimates the normalized probability of word $w_{i+j}$ appearing in the context of central word $w_{i}$ over all the words in the vocabulary. We employed a negative sampling method\cite{mikolov2013distributed} to compute this probability. We have explored the influence of different window sizes, i.e. $k$, on training time and the effect on outdated comments detection. The experimental results show that $k=2$, i.e. the window size is $5$, is the best choice with the relatively lower training time and the better effect on outdated comment detection, detailed in Appendix C.

To calculate the similarity between comment and code, we define three types of similarity measures.

\begin{enumerate}[(1)]
\item Word to word: Given two words $w_{1}$ and $w_{2}$, we define their semantic similarity as the cosine similarity between their learned word embeddings:
\begin{equation}
sim(w_{1},w_{2})=\cos (\mathbf{w}_{1} ,\mathbf{w}_{2} )=\frac{\mathbf{w}^{T}_{1} \mathbf{w}_{2} }{\parallel \mathbf{w}_{1} \parallel \parallel \mathbf{w}_{2} \parallel } \label{eq3}
\end{equation}
\item Word to sentence: Given a word $w$ and a sentence $S$, the similarity between them is computed as the maximum similarity between $w$ and any word $w^{\prime }$ in $S$:
\begin{equation}
sim(w,S)=\max_{w^{\prime }\in S} sim(w,w^{\prime })
\end{equation}
\item Sentence to sentence: Between two sentences $S_{1}$ and $S_{2}$, we define their semantic similarity as follows:
\begin{equation}
sim(S_{1},S_{2})=\frac{sim(S_{1}\rightarrow S_{2})+sim(S_{2}\rightarrow S_{1})}{2} 
\end{equation}
\noindent where
\begin{equation}
sim(S_{1}\rightarrow S_{2})=\frac{\sum_{w\in S_{1}} sim(w,S_{2})}{n} 
\end{equation}
\noindent where $n$ denotes the number of words in $S_{1}$.
\end{enumerate}

We considered the similarities between the comment and the code before and after the change. If the comment and the code have high similarity before the change and they have low similarity after the code change, then there is a high probability that the comment needs to be updated to ensure the consistency between code and comment. Similarly, we calculated the similarity between the comment and the changed statement before and after the change. In addition, we also compared the differences between the similarity of the comment and the changed statement before and after the change. 

We also extended the relation features based on our previous version. Inspired by word analysis in code features and comment features, if the number of words that the code and comment have in common before and after the change has changed significantly, then the comment is likely to be outdated. Therefore, we put the distance of token pairs the comment and code have in common before and after the change in the relation features. TABLE \ref{relationship} shows all the relation features.

\begin{table}[htb]
\centering
\caption{Relation features.}
\label{relationship}
\begin{tabular}{|l|l|}
\hline
\textbf{Feature} &
  \textbf{Description} \\ \hline
The distance of comment and changed statement similarity &
  \begin{tabular}[c]{@{}l@{}}The similarity of comment and changed statement before the change minus\\ the similarity of comment and changed statement after the change. The \\calculation method is shown in formula \ref{eqa}.\end{tabular} \\ \hline
The distance of comment token and code similarity &
  \begin{tabular}[c]{@{}l@{}}The average of the similarity of comment tokens and old code minus \\ the average of the similarity of comment tokens and new code. The \\ calculation method is shown in formula \ref{eqb}.\end{tabular} \\ \hline
The distance of comment and code similarity &
  \begin{tabular}[c]{@{}l@{}}The similarity of comment and old code minus the similarity of comment \\ and new code. The calculation method is shown in formula \ref{eqc}.\end{tabular} \\ \hline
The distance of token pairs comment and code have in common &
  \begin{tabular}[c]{@{}l@{}}The number of token pairs the comment and old code have in common \\minus the number of token pairs the comment and new code have \\ in common. \end{tabular} \\ \hline
\end{tabular}
\end{table}

\begin{equation}
D_{cmt\rightarrow smt}=|sim(S_{cmt},S_{smt})-sim(S_{cmt,}S_{smt^{\prime }})|\label{eqa}
\end{equation}
\begin{equation}
D_{token\rightarrow code}=\frac{\sum^{N}_{i} |sim(w^{i}_{cmt},S_{code})-sim(w^{i}_{cmt},S_{code^{\prime }})|}{N} \label{eqb}
\end{equation}
\begin{equation}
D_{cmt\rightarrow code}=|sim(S^{}_{cmt},S_{code})-sim(S_{comment},S_{code^{\prime }})|\label{eqc}
\end{equation}

\noindent where $S_{cmt}$ represents the comment, $S_{smt}$ represents the change statement before the change, $S_{smt^{\prime }}$ represents the change statement after the change, $w^{i}_{cmt}$ represents the $i$th token in the comment, $S_{code}$ represents the code before the change, $S_{code^{\prime }}$ represents the code after the change, and $N$ represents the number of tokens in the comment.

 The similarity feature calculated by the formula above is a decimal value between 0 and 1. In addition, we explored the impact of using the original feature values (decimal values) and artificially defined high and low similarity thresholds, on the outdated comment detection task (detailed in Appendix D).

\subsection{Machine learning algorithms}
In this section, we treat the problem of detecting outdated comments during code changes as a binary classification problem, with label "1" (comment outdated, positive sample) and label "0" (comment not outdated, negative sample) for binary classification. Whether a comment is outdated can be determined by code changes, comments, and the relation between comment and code. Based on the features above, we used random forest\cite{ho1995random} and other machine learning algorithms to classify outdated and not outdated comments.

The random forest\cite{ho1995random} algorithm constructs a large number of basic classifiers and lets them vote for the most likely category. In our case study, after cross-validation grid search (detailed in Appendix E), we selected 200 n\_estimators and set the random subset value of features as 'sqrt', used gini as the criterion, gini score is calculated as follows:

\begin{equation}
Gini(D)=1-(\frac{|c_0 |}{|D|} )^2-(\frac{|c_1 |}{|D|} )^2\label{eq}
\end{equation}

\noindent where $D$ is the sample set, $|D|$ is the number of samples in set $D$, $|c_0|$ is the number of negative samples in $D$, and $|c_1|$ is the number of positive samples in $D$

The gain of a classifier is calculated as follows:

\begin{equation}
Gain(D,A)=Gini(D)-\frac{|D_1 |}{|D|}Gini(D_1 )-
\frac{|D_2 |}{|D|}Gini(D_2 )\label{eq}
\end{equation}

\noindent where $A$ is an attribute, which is divided into two subsets $D_1$ and $D_2$, $|D_1 |$ is the number of samples in set $D_1$ and $|D_2 |$ is the number of samples in set $D_2$.

Compared to the previous version, we compared the performance of different learning-based models: XGBoost\cite{2016XGBoost} is a scalable end-to-end tree enhancement system, an optimized distributed gradient enhancement library designed to be efficient, flexible, and portable. Logistic regression\cite{dreiseitl2002logistic}, is a generalized linear regression analysis model, which belongs to supervised learning in machine learning and is often used to solve dichotomous problems. Naive bayes is a series of simple probabilistic classifiers based on the assumption of strong (plain) independence between features using Bayes' theorem. This classifier model assigns class labels to problem instances expressed in terms of feature values, and class labels are taken from a finite set. It is not a single algorithm for training such a classifier, but a series of algorithms based on the same principle: all plain bayesian classifiers assume that each feature of the sample is uncorrelated with every other feature. SVM\cite{cortes1995support} is a class of generalized linear classifiers that binary classifies data in a supervised learning fashion, with a decision boundary of the maximum margin hyperplane solved for the learned samples. Decision tree\cite{quinlan1987simplifying} is a tree structure. Each non-leaf node represents a test on a feature attribute, each branch represents the output of the feature attribute on a value domain, and each leaf node holds a category. The process of decision-making using a decision tree\cite{quinlan1987simplifying} starts at the root node, tests the corresponding feature attribute of the item to be classified, and selects the output branch according to its value until it reaches the leaf node, where the category stored in the leaf node is used as the decision result.

To evaluate the performance of different learning-based models, we referred to the model reliability curve\cite{pedregosa2011scikit,domingos1996beyond,zadrozny2001obtaining}, which is a probabilistic model evaluation metric for algorithms. It is a curve with the predicted label value as the horizontal coordinate and the true label value as the vertical coordinate.
Therefore when we draw the reliability curve closer to the diagonal, we consider that the performance of this model learner is better.

\section{EXPERIMENTS setups}
\subsection{Data collection}
Our study focuses on detecting the outdated comments in method-type code-comment pairs and block-type code-comment pairs, and we collected the data from open-source projects, which are shown in TABLEs \ref{dataset1} and \ref{dataset}. All the projects can be found in our replication package.

\begin{table}[htb]
\centering
\caption{Method-type and block-type code-comment pairs used in the experiments.}
\label{dataset1}
\begin{tabular}{|l|l|l|l|ll|}
\hline
\multirow{2}{*}{\textbf{Type}} & \multirow{2}{*}{\textbf{Commit class}} & \multirow{2}{*}{\textbf{Code-comment pair}} & \multirow{2}{*}{\textbf{Change}} & \multicolumn{2}{l|}{\textbf{Comment}} \\ \cline{5-6} 
            &       &       &       & \multicolumn{1}{l|}{\textbf{Changed}} & \textbf{Unchanged} \\ \hline
Method-type & -     & 27,665 & 48,097 & \multicolumn{1}{l|}{8.01\%}  & 91.99\%   \\ \hline
Block-type  & -     & 21,983 & 35,819 & \multicolumn{1}{l|}{8.79\%}  & 91.21\%   \\ \hline
Total       & 40,247 & 49,648 & 83,916 & \multicolumn{1}{l|}{16.80}   & 83.20\%   \\ \hline
\end{tabular}
\end{table}

\begin{table}[htb]
\centering
\caption{Projects used in the experiments.}
\label{dataset}
\begin{tabular}{|l|l|l|l|ll|}
\hline
\multirow{2}{*}{\textbf{Project}} & \multirow{2}{*}{\textbf{Commit class}} & \multirow{2}{*}{\textbf{Code-comment pair}} & \multirow{2}{*}{\textbf{Change}} & \multicolumn{2}{l|}{\textbf{Comment}}           \\ \cline{5-6} 
          &       &       &       & \multicolumn{1}{l|}{\textbf{Changed}} & \textbf{Unchanged} \\ \hline
dcm4che   & 2,113  & 2,191  & 3,597  & \multicolumn{1}{l|}{21.77\%} & 78.22\%   \\ \hline
Ejbca     & 2,508  & 2,532  & 4,655  & \multicolumn{1}{l|}{10.62\%} & 89.38\%   \\ \hline
freecol   & 2,230  & 1,830  & 3,374  & \multicolumn{1}{l|}{7.05\%}  & 92.95\%   \\ \hline
ghidra    & 2,846  & 5,072  & 6,136  & \multicolumn{1}{l|}{67.59\%} & 32.41\%   \\ \hline
greenDAO  & 625   & 773   & 1,055  & \multicolumn{1}{l|}{6.34\%}  & 93.66\%   \\ \hline
Guice     & 1,401  & 1,422  & 2,024  & \multicolumn{1}{l|}{3.16\%}  & 96.84\%   \\ \hline
Hieos     & 547   & 1,254  & 2,610  & \multicolumn{1}{l|}{12.92\%} & 87.08\%   \\ \hline
Hsqldb    & 1,139  & 1,505  & 2,592  & \multicolumn{1}{l|}{7.44\%}  & 92.56\%   \\ \hline
Htmlunit  & 1,392  & 1,287  & 1,432  & \multicolumn{1}{l|}{5.05\%}  & 94.95\%   \\ \hline
J2objc    & 3,833  & 3,453  & 6,007  & \multicolumn{1}{l|}{5.30\%}  & 94.70\%   \\ \hline
Jamwiki   & 899   & 980   & 1,772  & \multicolumn{1}{l|}{22.55\%} & 77.44\%   \\ \hline
Jedit     & 1,119  & 1,129  & 2,214  & \multicolumn{1}{l|}{13.82\%} & 86.18\%   \\ \hline
Joda-time & 547   & 797   & 994   & \multicolumn{1}{l|}{62.74\%} & 37.26\%   \\ \hline
Kablink   & 1,983  & 2,031  & 3,833  & \multicolumn{1}{l|}{25.06\%} & 74.94\%   \\ \hline
Makagiga  & 1,593  & 1,762  & 2,745  & \multicolumn{1}{l|}{8.06\%}  & 91.94\%   \\ \hline
Neo4j     & 2,100  & 2,613  & 3,796  & \multicolumn{1}{l|}{4.29\%}  & 95.71\%   \\ \hline
Omegat    & 837   & 906   & 1,534  & \multicolumn{1}{l|}{5.74\%}  & 94.26\%   \\ \hline
Opennms   & 1,269  & 923   & 1,680  & \multicolumn{1}{l|}{8.67\%}  & 91.33\%   \\ \hline
Realm-Java               & 5,957                          & 10698                  & 20,592                   & \multicolumn{1}{l|}{12.11\%} & 87.89\% \\ \hline
Titan     & 3,338  & 3,740  & 6,502  & \multicolumn{1}{l|}{3.40\%}  & 96.60\%   \\ \hline
Txm       & 778   & 851   & 1,397  & \multicolumn{1}{l|}{14.57\%} & 85.43\%   \\ \hline
Zeppelin  & 2,585  & 1,899  & 3,375  & \multicolumn{1}{l|}{5.53\%}  & 94.47\%   \\ \hline
Total     & 40,247 & 49,648 & 83,916 & \multicolumn{1}{l|}{16.80}   & 83.20\%   \\ \hline
\end{tabular}
\end{table}

These projects score over 4.5 out of 5 in SourceForge\footnote{\url{https://sourceforge.net/}} or GitHub\footnote{\url{https://github.com/}} and are widely used for source code analysis and research \cite{huang2020towards,Huanchao2019}. They have different types of application domains, including text editors, management systems, collaboration software, wiki engines, 2D graphics frameworks, and games.

From these projects, we collected 49,648 code-comment pairs from 40,247 commits. Among these pairs, the number of method-type code-comment pairs is 27,665 and the number of block-type code-comment pairs is 21,983. Positive samples accounted for 16.8\%, and negative samples accounted for 83.2\%. Before training, we split data into the training set and test set at a ratio of 7:3, and the best hyperparameters of the model have been obtained by grid search and cross-validation (detailed in Appendix E) on the training set, and finally tested on the test set.

\subsection{Research questions}

The main goal of our approach is to automatically detect outdated comments based on code changes. Therefore, we focus on the effectiveness, applicability, and practicability of CoCC. We mainly focused on the following specific issues. The first is whether different machine learning models have an impact on the performance of CoCC. We selected 6 machine learning methods for comparison. The second is the effectiveness of our model in predicting outdated comments. We compared different baselines to CoCC. The third is the contribution of different features to the effectiveness of detecting outdated comments. In addition, we try to find a minimum effective feature set to detect outdated comments. The fourth is the applicability to other languages of CoCC. We extended the experiment to Python. Finally, we pay attention to the practicality of CoCC. We used trained CoCC to detect outdated comments in the latest commits of the open-source projects, and invited the programmers to manually check the outdated comments detected by CoCC. The experimental results prove that CoCC can detect outdated comments in the actual scene.

Therefore, we mainly focus on the following research questions:
\begin{itemize}
\item RQ1: Which learning-based model is suitable for the outdated comment detection task?
\item RQ2: How effective is the model in predicting outdated comments?
\item RQ3: How much does each feature contribute to the effectiveness of the prediction?
\item RQ4: How effective is CoCC to detect outdated comments in other programming languages?
\item RQ5: Can CoCC detect outdated comments in the latest commits?
\end{itemize}

\subsection{Evaluation criterion}

In this study, we used three evaluation scores, including precision, recall, and f1 score, to evaluate the performance of our models in outdated comments detection. Precision is the number of samples correctly predicted by the model as a percentage of the total test sample. Recall is the percentage of outdated comments found by the model in the total sample of outdated comments in the test set, and f1 score is a metric used to measure the accuracy of learning-based models in statistics. It considers both the precision and recall of the classification model. Positive samples are code-comment pairs with outdated comments, negative samples are code-comment pairs without outdated comments, TP represents true positive samples, FP represents false positive samples, TN represents true negative samples, FN represents false negative samples, then, the precision, recall, and f1 score are calculated as follows:

\begin{equation}
Precision = \frac{TP}{TP+FP}\label{eq}
\end{equation}

\begin{equation}
Recall = \frac{TP}{TP+FN}\label{eq}
\end{equation}

\begin{equation}
F1 = 2 * \frac{Precision*Recall}{Precision+Recall} \label{eq}
\end{equation}

\section{Results}
\label{result}

\subsection*{RQ1: Which learning-based model is suitable for the outdated comment detection task?}

We compared the performance of 6 different models: random forest, XGBoost, logistic regression, naive bayes, SVM, and decision tree.

\begin{figure}[htbp]
\centering
\includegraphics[scale=0.2]{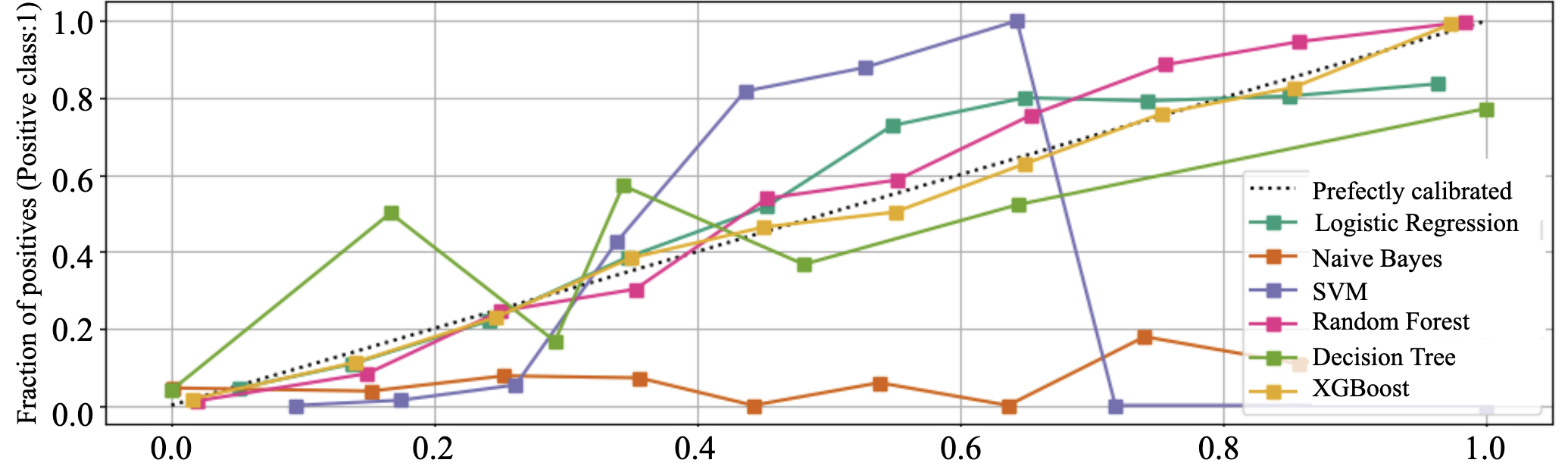}
\caption{Calibration curves and the distribution of predicted probabilities.}
\label{calibration}
\end{figure}

To achieve the best performance for different models, we use grid search and cross-validation to obtain the best hyperparameter settings for each model (detailed in Appendix E). Then, we trained each model with the training dataset and then plotted a calibration curve\cite{pedregosa2011scikit,domingos1996beyond,zadrozny2001obtaining} (also called a reliability plot) using the predicted probabilities from the test dataset. It is a curve with the predicted label value as the horizontal coordinate and the true label value as the vertical coordinate. The experiment results are shown in FIGURE \ref{calibration} and TABLE \ref{classfi}.

\begin{table}[htb]
\centering
\caption{Result for different classifiers.}
\label{classfi}
\begin{tabular}{|l|l|l|l|}
\hline
\textbf{Model}               & \textbf{Precison}        & \textbf{Recall}          & \textbf{F1}             \\ \hline
Logistic regression & 79.3\%          & 53.5\%          & 0.639          \\ \hline
Naive bayes         & 18.4\%          & \textbf{97.1\%} & 0.310          \\ \hline
SVM                 & 86.2\%          & 63.5\%          & 0.731          \\ \hline
Random forest       & \textbf{92.1\%} & 78.9\%          & \textbf{0.850} \\ \hline
Decision tree       & 81.1\%          & 76.2\%          & 0.786          \\ \hline
XGBoost             & 89.2\%          & 77.1\%          & 0.827          \\ \hline
\end{tabular}
\end{table}

From FIGURE \ref{calibration}, the calibration curves of random forest, XGBoost trained by our data are more towards the diagonal. From TABLE \ref{classfi}, the comprehensive performance of random forest and XGBoost are better than other models, so they are more suitable as learning-based models for our task. We concluded that the random forest model and XGBoost perform better on the feature-based outdated comment detection task, so for the research questions we selected random forest and XGBoost as the base models.

\subsection*{RQ2: How effective is the model in predicting outdated comments?}

To answer this question, we selected four baselines to compare with our model. Because there were not many previous studies on outdated comment detection, we also added a rule-based baseline. The following is the baseline used for comparison in the experiment: the first is our previous version\cite{liu2018automatic}, where CoCC expanded its features and deleted some highly relevant features; The second is the random guess, that is, in theory, half of the correct rate detects outdated comments; The third is a rule-based approach, if the difference of code similarity before and after the change is more than 5\%, it will be considered as outdated comment. We iterated different values in the step of 5\%, of which 5\% is the best. The fourth baseline is OCD\cite{liu2021just}, which is a neural-network-based outdated comment detection model proposed by Liu et al. since OCD only focuses on method-type outdated comment detection, we only compare our method-type data with it.

The data used in the experiment is shown in Section 3.1. For OCD, since it only focuses on the outdated detection of method-type comments, we only take out the method-type data in TABLE \ref{dataset1} to compare with it. The data comes from different projects, so the features may have different ranges, we normalize the features across all projects before training, while not one project at a time: for discrete features and binary features (such as whether include the return value, 0 or 1), we do not deal with them; For numerical continuous features (such as the similarity), we use formula \ref{standardize} to standardize the features.

\begin{equation}
{}f_{0}=\frac{f_{0}-\overline{f} }{\sigma (f)} \label{standardize}
\end{equation}

\noindent where $f_{0}$ represents a feature, $\overline{f}$ represents the average value of the feature, and $\sigma (f)$ represents the standard deviation of the feature.

\begin{table}[htb]
\centering
\caption{Results of the different outdated comment detection methods.}
\label{resultrq2}
\begin{tabular}{|ll|l|l|l|l|}
\hline
\multicolumn{2}{|l|}{\textbf{Method}}                                & \textbf{Comment type}    & \textbf{Precision}       & \textbf{Recall}          & \textbf{F1}             \\ \hline
\multicolumn{2}{|l|}{\multirow{3}{*}{Previous version}} & Method          & 83.4\%          & 61.0\%          & 0.704          \\ \cline{3-6} 
\multicolumn{2}{|l|}{}                                      & Block           & 89.6\%          & 87.6\%          & 0.886          \\ \cline{3-6} 
\multicolumn{2}{|l|}{}                                      & Method \& block & 88.4\%          & 75.9\%          & 0.817          \\ \hline
\multicolumn{2}{|l|}{\multirow{3}{*}{Random guess}}    & Method          & 50\%            & 50\%            & 0.5            \\ \cline{3-6} 
\multicolumn{2}{|l|}{}                                      & Block           & 50\%            & 50\%            & 0.5            \\ \cline{3-6} 
\multicolumn{2}{|l|}{}                                      & Method \& block & 50\%            & 50\%            & 0.5            \\ \hline
\multicolumn{2}{|l|}{\multirow{3}{*}{Rule based}}           & Method          & 69.4\%          & 29.8\%          & 0.417          \\ \cline{3-6} 
\multicolumn{2}{|l|}{}                                      & Block           & 60.0\%          & 29.4\%          & 0.395          \\ \cline{3-6} 
\multicolumn{2}{|l|}{}                                      & Method \& block & 64.1\%          & 29.6\%          & 0.405          \\ \hline
\multicolumn{2}{|l|}{\multirow{3}{*}{OCD}}                  & Method          & 80.5\%          & 12.5\%          & 0.216          \\ \cline{3-6} 
\multicolumn{2}{|l|}{}                                      & Block           & -               & -               & -              \\ \cline{3-6} 
\multicolumn{2}{|l|}{}                                      & Method \& block & -               & -               & -              \\ \hline
\multicolumn{2}{|l|}{\multirow{3}{*}{\textbf{CoCC (random forest)}}} & Method & \textbf{89.1\%} & 67.1\%          & \textbf{0.765} \\ \cline{3-6} 
\multicolumn{2}{|l|}{}                                      & Block           & \textbf{93.5\%} & 88.5\%          & \textbf{0.909} \\ \cline{3-6} 
\multicolumn{2}{|l|}{}                                      & Method \& block & \textbf{92.1\%} & \textbf{78.9\%} & \textbf{0.850} \\ \hline
\multicolumn{2}{|l|}{\multirow{3}{*}{\textbf{CoCC (XGBoost)}}}       & Method & 83.9\%          & \textbf{67.8\%} & 0.750          \\ \cline{3-6} 
\multicolumn{2}{|l|}{}                                      & Block           & 92.8\%          & \textbf{88.9\%} & 0.908          \\ \cline{3-6} 
\multicolumn{2}{|l|}{}                                      & Method \& block & 89.2\%          & 77.1\%          & 0.827          \\ \hline
\end{tabular}
\end{table}

The experiment results are shown in TABLE \ref{resultrq2}, CoCC is superior to baseline on all evaluation scores. For method-type outdated comment detection, random forest CoCC performs best on precision and f1, and XGBoost CoCC performs best on recall. For block-type outdated comment detection, random forest CoCC performs best on precision and f1, and XGBoost CoCC performs best on recall; For the mixed two types of outdated comment detection, the random forest CoCC performs best on all evaluation scores.

\subsection*{RQ3: How much does each feature contribute to the effectiveness of the prediction?}

To explore the factors affecting prediction effectiveness, we further investigated the contribution of different features. We took out the code features, the comment features, and the relation features to train the random forest and XGBoost classifiers, separately. The experiment results are shown in TABLE \ref{resultrq3}.

\begin{table}[htb]
\centering
\caption{Results of different features.}
\label{resultrq3}
\begin{tabular}{|l|l|l|l|l|}
\hline
\textbf{CoCC} & \textbf{Feature}          & \textbf{Precision}       & \textbf{Recall}          & \textbf{F1}             \\ \hline
\multirow{4}{*}{Random forest} & Code feature & 80.2\%          & \textbf{71.9\%} & \textbf{0.758} \\ \cline{2-5} 
     & Comment feature  & 74.4\%          & \textbf{59.5\%} & \textbf{0.661} \\ \cline{2-5} 
     & Relation feature & 84.2\%          & \textbf{66.8\%} & \textbf{0.745} \\ \cline{2-5} 
     & All feature      & \textbf{92.1\%} & \textbf{78.9\%} & \textbf{0.850} \\ \hline
\multirow{4}{*}{XGBoost}       & Code feature & \textbf{82.8\%} & 64.4\%          & 0.725          \\ \cline{2-5} 
     & Comment feature  & \textbf{78.9\%} & 53.9\%          & 0.641          \\ \cline{2-5} 
     & Relation feature & \textbf{85.2\%} & 63.3\%          & 0.726          \\ \cline{2-5} 
     & All feature      & 89.2\%          & 77.1\%          & 0.827          \\ \hline
\end{tabular}
\end{table}

From  TABLE \ref{resultrq3}, first, whether it is CoCC based on random forest or CoCC based on XGBoost, the performance of code features and relation features should be better than that of comment features. Second, code features, comment features, and relation features are not as good as their combined performance. This also shows the effectiveness of the features we selected.

In addition, we explored the effective contribution of a single feature. We expect to find a minimum effective feature set to detect outdated comments. Therefore, we first need to calculate the importance of each feature, which needs the help of the random forest. The random forest is a set of decision trees. Each decision tree is a set of internal nodes and leaves. In the internal node, the selected feature is used to make a decision on how to divide the data set into two separate sets with similar responses within. The features for internal nodes are selected with some criterion, which for our task tasks is gini impurity gain. We can measure how each feature decreases the impurity of the split (the feature with the highest decrease is selected for the internal node). For each feature, we can collect how on average it decreases the impurity. The average over all trees in the forest is the measure of the feature importance. Specifically, we count the characteristic importance score as $FIS$ (feature importance score), assuming there are $J$ characteristics, $I$ decision tree, $C$ categories, then the gini score of node $q$ of the $i$th tree is calculated as formula \ref{impgini}:

\begin{equation}
Gini^{(i)}_{q}=\sum^{|C|}_{c=1} \sum^{}_{c^{\prime }\neq c} p^{(i)}_{qc}p^{(i)}_{qc^{\prime }}=1-\sum^{|C|}_{c=1} (p^{(i)}_{qc})^{2}\label{impgini}
\end{equation}

where, $C$ indicates that there are $C$ categories (here $C$ is $2$, comment outdated or not), and $p_{qc}$ indicates the proportion of category $c$ in node $q$. The importance of node $q$ (feature $X_{j}$) in the $i$th tree, that is, the change of gini score before and after node $q$ is:

\begin{equation}
FIS^{Gini(i)}_{jq}=Gini^{(i)}_{q}-Gini^{(i)}_{l}-Gini^{(i)}_{r}
\label{eq}
\end{equation}

where $Gini^{(i)}_{l}$ and $Gini^{(i)}_{r}$ represent gini score of two new nodes after branching. If the node of feature $X_{j}$ in decision tree $i$ is set $Q$, then the importance of $X_{j}$ in the $i$th tree is:

\begin{equation}
FIS^{Gini(i)}_{j}=\sum^{}_{q\in Q} FIS^{Gini(i)}_{jq}
\label{eq}
\end{equation}

Then the sum of the importance of feature $X_{j}$ on all trees in the random forest is:
\begin{equation}
FIS^{Gini}_{j}=\sum^{I}_{i=1} FIS^{Gini(i)}_{j}\label{eq}
\end{equation}

Finally, normalize the sum to get the importance score of the feature $X_{j}$:
\begin{equation}
FIS^{Gini}_{j}=\frac{FIS^{Gini}_{j}}{\sum^{J}_{j^{\prime }=1} FIS^{Gini}_{j^{\prime }}} \label{eq}
\end{equation}

We use the above method to calculate the importance of features. TABLE \ref{topf} shows the 15 most important features for outdated comment detection. In addition, we use these 15 features to detect outdated comments, and the results are shown in TABLE \ref{topres}.

\begin{table}[htb]
\centering
\caption{Minimum effective feature sets.}
\label{topf}
\begin{tabular}{|l|}
\hline
\textbf{Feature name}                                \\ \hline
The distance of comment and code similarity                   \\ \hline
The distance of comment token and code similarity             \\ \hline
The distance of comment and changed statement similarity      \\ \hline
The proportion of code lines in code-comment pair                 \\ \hline
The proportion of changed statement lines in code-comment pair    \\ \hline
Word analysis                                                 \\ \hline
The distance of token pairs comment and code have in common \\ \hline
Method invocation update                                      \\ \hline
Number of changes                                             \\ \hline
Contain return or not                                        \\ \hline
Variable declaration update                                   \\ \hline
Method invocation delete                                      \\ \hline
Method renaming                                               \\ \hline
Return type                                                   \\ \hline
Parameter renaming                                            \\ \hline
\end{tabular}
\end{table}

\begin{table}[htb]
\centering
\caption{Result of minimum effective feature sets.}
\label{topres}
\begin{tabular}{|l|l|l|l|}
\hline
\textbf{Feature}                        & \textbf{Precision} & \textbf{Recall} & \textbf{F1}    \\ \hline
Minimum effective feature sets & 89.4\%    & 75.2\% & 0.817 \\ \hline
All feature                    & 92.1\%    & 78.9\% & 0.850 \\ \hline
\end{tabular}
\end{table}

From TABLE \ref{topf}, we can reach the following key conclusions: first, the semantic similarity features of code comments are very important for the detection of outdated comments. The inspiration is that if the code is modified and the similarity between the codes and comments decreases after the change, the comments may be outdated. Second, the difference of token pairs comment and code have in common before and after the change affects whether the comment should be updated, the inspiration is that when the code change involves the same tokens used in code and comment, the comment needs to be updated in time to keep the consistency between the comments and the codes. Third, changes related to the method declaration affect whether the comment is outdated. The inspiration is that if the programmer modifies the name, parameter, or return type of the method, and the comment contains relevant descriptions of the method name, parameter, or return type, then when the code change occurs, the comment needs to be updated in time to ensure the consistency between code and comment.

From TABLE \ref{topres}, although the minimum effective feature set performed well on the outdated comment detection task, there is still a certain gap between its performance and all features, which shows the effectiveness of all features we selected.

\subsection*{RQ4: How effective is CoCC to detect outdated comments in other programming languages?}

CoCC is a method for detecting outdated comments based on code changes. For a programming language, we extract code features, comment features, and relation features to determine whether the comments are outdated. We have verified the effectiveness of CoCC on Java in RQ2. In addition, we want to explore whether the features selected by CoCC are applicable to other programming languages. The tool used to collect code-comment pairs and extract features in our method is ChangeDistiller. At present, it only supports Java language, so if other languages have available tools for us to collect and label data automatically, we can collect enough data in other languages for the experiment. 

To verify the applicability to other languages of CoCC, we chose Python as the second research language. Unfortunately, Python does not have a tool like ChangeDistiller for us to collect data and extract 48 of the features such as refactoring, class attribute change, and method declaration. We manually collected and labeled 5000 samples for the experiment. The Python data comes from open-source projects with high stars on GitHub, detailed in TABLE \ref{pydata}. The experimental results are shown in TABLE \ref{pyresult}:

\begin{table}[htb]
\centering
\caption{Python data source.}
\label{pydata}
\begin{tabular}{|l|l|}
\hline
\textbf{Project} & \textbf{GitHub link}               \\ \hline
Django           & https://github.com/django/django   \\ \hline
Sanic            & https://github.com/sanic-org/sanic \\ \hline
Pipenv           & https://github.com/pypa/pipenv     \\ \hline
\end{tabular}
\end{table}

\begin{table}[htb]
\centering
\caption{Result of the experiment on Python data.}
\label{pyresult}
\begin{tabular}{|l|l|l|l|}
\hline
\textbf{Programming language} & \textbf{Precision} & \textbf{Recall} & \textbf{F1}    \\ \hline
Python dataset       & 89.4\%    & 75.7\% & 0.820 \\ \hline
Java dataset         & 92.1\%    & 78.9\% & 0.850 \\ \hline
\end{tabular}
\end{table}

From TABLE \ref{pyresult}, CoCC performs slightly worse on Python than on Java but still has higher precision, recall, and f1. In addition, due to the similarity of object-oriented programming languages, CoCC can be applied to different programming languages. Due to the lack of automated tools to collect Python data, we manually collected the data, which results in the lack of training data.  We considered it as one of the threats to validity, detailed in threat to validity.

Of the code features, comment features, and relation features, only the code features are related to the programming languages, and because of the similarity of object-oriented programming languages, programmers also have similar coding styles and constraints when using different programming languages, such as class design methods, so in general, our methods are applicable to different programming languages, we can use similar methods to detect outdated comments in different programming languages.

\subsection*{RQ5: Can CoCC detect outdated comments in the latest commits?}

To answer this question, we used trained CoCC to detect the outdated comments in the latest commits of researched project and feedback to the developers. We got the programmers' email, Twitter, and other contact information from GitHub. We contacted more than 50 developers, such as primetomas in EJBCA, rangerRick in FreeCol, and deejgregor in OpenNMS, unfortunately, we didn't receive any reply within six weeks, as a result, we invited 20 programmers with 5 years or more of development experience for manual check in the form of the questionnaire. All the participants come from internet companies in China, such as Alibaba and Tencent, etc., and all the participants were graduates of our research laboratory, but none were authors of this paper.

\begin{figure}[htbp]
\centering
\includegraphics[scale=0.20]{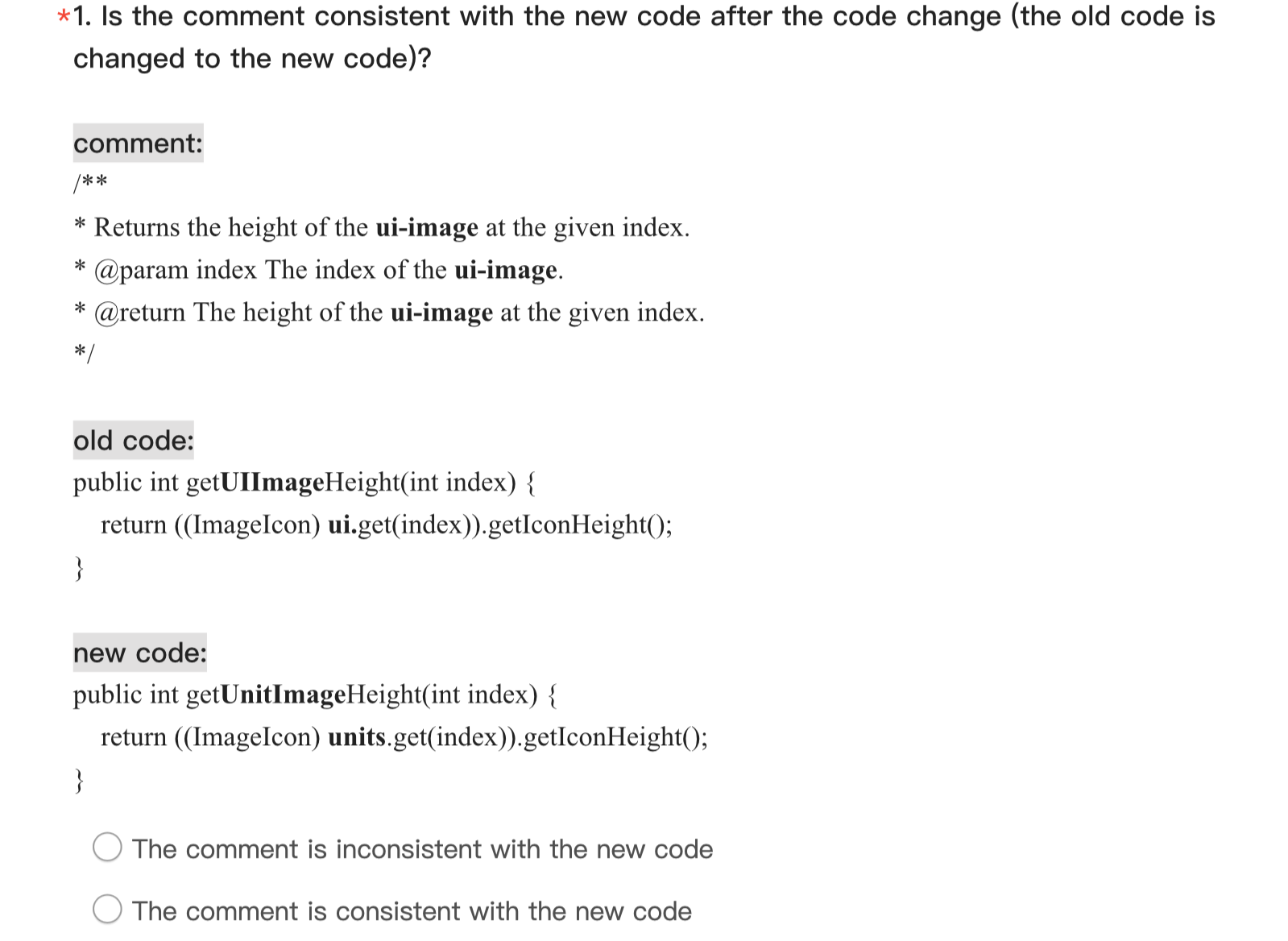}
\caption{A sample in the questionnaire.}
\label{question}
\end{figure}

We selected 300 latest commits from high-star open-source Java projects EJBCA, Freecol, and OpenNMS respectively, and used our trained model to detect outdated comments. Finally, 39 outdated comments were detected from 900 commits, 14 of them are repetitive, and we took out 25 that are not repeated as questionnaire samples. The sample in the questionnaire is shown in FIGURE \ref{question}.

In FIGURE \ref{question}, the “old code” represents the code before the change, and the “new code” represents the code after the change. Question is to check whether the "comment" is consistent with the "new code" after the code change. The options are set as follows: (1) The comment is inconsistent with the new code. (2) The comment is consistent with the new code.

The survey results are shown in TABLE \ref{rq4}, we received 468 answers to option (1) and 32 answers to option (2) from 20 questionnaires ($20\times 25=500$, total answers). That is, of the 25 outdated comments in the latest version detected by CoCC, 93.6\% of comments programmers think that they are outdated comments, and 6.4\% of comments programmers think that they are not outdated.

\begin{table}[htb]
\centering
\caption{Result from questionnaires.}
\label{rq4}
\begin{tabular}{|l|l|l|}
\hline
\textbf{Option}                                & \textbf{Number} & \textbf{Percentage} \\ \hline
The comment is inconsistent with the new code. & 468             & 93.6\%              \\ \hline
The comment is consistent with the new code.   & 32              & 6.4\%               \\ \hline
\end{tabular}
\end{table}

If less than or equal to 3 programmers think that a comment is not outdated, then the comment is considered to be outdated. Of the 25 outdated comments detected by CoCC, 23 are outdated, which also proves that CoCC can help programmers find outdated comments in actual development, urge programmers to update comments in time, and maintain the consistency between code and comment.

At the end of the questionnaire, we also left a question: if there is a real-time outdated comment detection tool available, would you like to use it? All participants in the survey have chosen to use it if the detection is accurate enough.

\section{RELATED WORK}
Source code comments play a critical role in program understanding. A survey of software maintainers conducted by De SouzaMastropaolo et al.\cite{mastropaolo2021empirical} found that developers use comments as a key element in understanding source code. It highlights the critical role of comments in software development and maintenance. Arafat et al.\cite{2009The} researched the density of comments in open-source software code to understand how and why open-source software remains high quality and maintains that quality at larger project scales. They found that successful open-source projects follow a consistent practice of documenting their source code. Furthermore, their findings show that comment density is independent of team and project.

Comments are considered documented knowledge for developers, and comment quality is important in evaluating software quality. McBurney et al.\cite{mcburney2016empirical} conducted an empirical study examining method-type comments of source code written by authors, readers, and automatic source code summarization tools. Their work discovered that the accuracy of human-written method-type comments could be estimated by the textual similarity of that method-type comments to the source code, addressing that good comments written by developers should have a high semantic similarity to the source code. Steidl et al.\cite{steidl2013quality} presented an approach for comment quality analysis and assessment, which was based on different comment categories using machine learning on Java and C/C++ programs. Their comprehensive quality model comprised quality attributes for each comment category based on four criteria: consistency throughout the project, completeness of the system documentation, coherence with the source code, and usefulness to the reader. Both aforementioned studies used one aspect of the textual similarity to determine the inconsistency between code and comment, which is insufficient for such a complicated problem. Aman et al.\cite{aman2015empirical} conducted an empirical analysis on the usefulness of local variable names and comments in software quality assessments from the perspective of six popular open-source software products. Their study showed that a method having longer-named local variables is more change-prone. To understand if and how database-related statements are commented in source code, Linares-Vasquez et al.\cite{linares2015developers} mined Java open-source projects that use JDBC for the data access layer from Github. They found that 77$\%$ of 33K+ source code methods do not have header comments. They also pointed out that existing comments were rarely updated when the related source code was modified.

Considering the importance of comments in programming practice, and the fact that outdated comments may confuse and mislead programmers, several researchers\cite{ibrahim2012relationship,fluri2007code,fluri2009analyzing} have investigated how code and comment co-evolve. Fluri et al.\cite{fluri2007code} conducted an empirical survey on three open-source systems to study how comments and source code co-evolved over time. Their investigation results showed that 97$\%$ of comment changes are done in the same revision as the associated source code change. In Fluri’s other study\cite{fluri2009analyzing}, eight different software systems were analyzed, with the finding that code and comment co-evolved in 90$\%$ of the cases in six out of the eight systems. Ibrahim et al.\cite{ibrahim2012relationship} believed that bug prediction models play important roles in the prioritization of testing and code inspection efforts. They studied comment update practices in three large open-source systems written in C and Java and found that a change in which a function and its comments are suddenly updated inconsistently, has a high probability of introducing a bug.

Doc comments are important for understanding an application programming interface (API), outdated Javadoc comments can mislead method callers. Tan et al.\cite{tan2012tcomment} proposed a tool called @tComment to test for outdated Javadoc comments. @tComment employs the Randoop tool to test whether the method properties (regarding null values and related exceptions) violate some constraints contained in Javadoc comments. Khamis et al.\cite{khamis2010automatic} examined the correlation between code quality and Javadoc comment code inconsistencies. However, only some simple issues were checked, e.g., whether the parameter names, return types, and exceptions in the @param, @return, and @throws tags were consistent with the actual parameter names, return types, and exceptions in the method, respectively. Their automatic comment analysis technique achieved a high accuracy largely because Javadoc comments are well structured, with few paraphrases and variants.

Malik et al.\cite{malik2008understanding} conducted a large empirical study to better understand the rationale for updating comments in four large open-source projects written in C. They investigated the rationale for updating comments along three dimensions: features of the changed function, features of the change itself, and time and code ownership features. Unlike our method, these two studies both used a method/function as the unit in detecting consistencies. TODO comments, which are used by developers to denote pending tasks, may lead to out-of-date comments when developers perform the mentioned tasks and then forget to remove the comments. Sridhara\cite{Sridhara2016Automatically} presented a novel technique to automatically detect the status of TODO comments using three aspects, including information retrieval, linguistics, and semantics. The results showed that his status checker achieved high accuracy, precision, and recall in checking whether a TODO comment was up-to-date.

Liu et al. used the neural network classification method called OCD\cite{liu2021just}, which they combined with their novel common attention mechanism to get the relation between code changes and comments from a large number of code submissions, so as to detect outdated comments. OCD takes a code change and an associated old comment as inputs and outputs the probability that this comment should be updated with this code change. Their attention mechanism can learn to effectively focus on and select the information that is important for outdated comment prediction.

Iammarino et al. presented an approach, based on topic modeling, for analyzing the comments' consistency to the source code\cite{iammarino2020topic}. A model was provided to analyze the quality of comments in terms of consistency since comments should be consistent with the source code they refer to. The results show a similarity in the trend of topic distribution and it emerges that almost all classes are associated with no more than 3 topics.

Stulova et al. proposed a technique and a tool, upDoc, to automatically detect code-comment inconsistency during code evolution\cite{stulova2020towards}. They built a map between the code and its documentation, ensuring that changes in the code match the changes in respective documentation parts. They conducted a preliminary evaluation using inconsistent examples from an existing dataset of Java open-source projects, showing that upDos can successfully detect them. They presented a roadmap for the further development of the technique and its evaluation.

\section{THREAT TO VALIDITY}

Several threats may affect the effectiveness of the experiment.

\begin{figure}[htbp]
\centering
\includegraphics[scale=0.15]{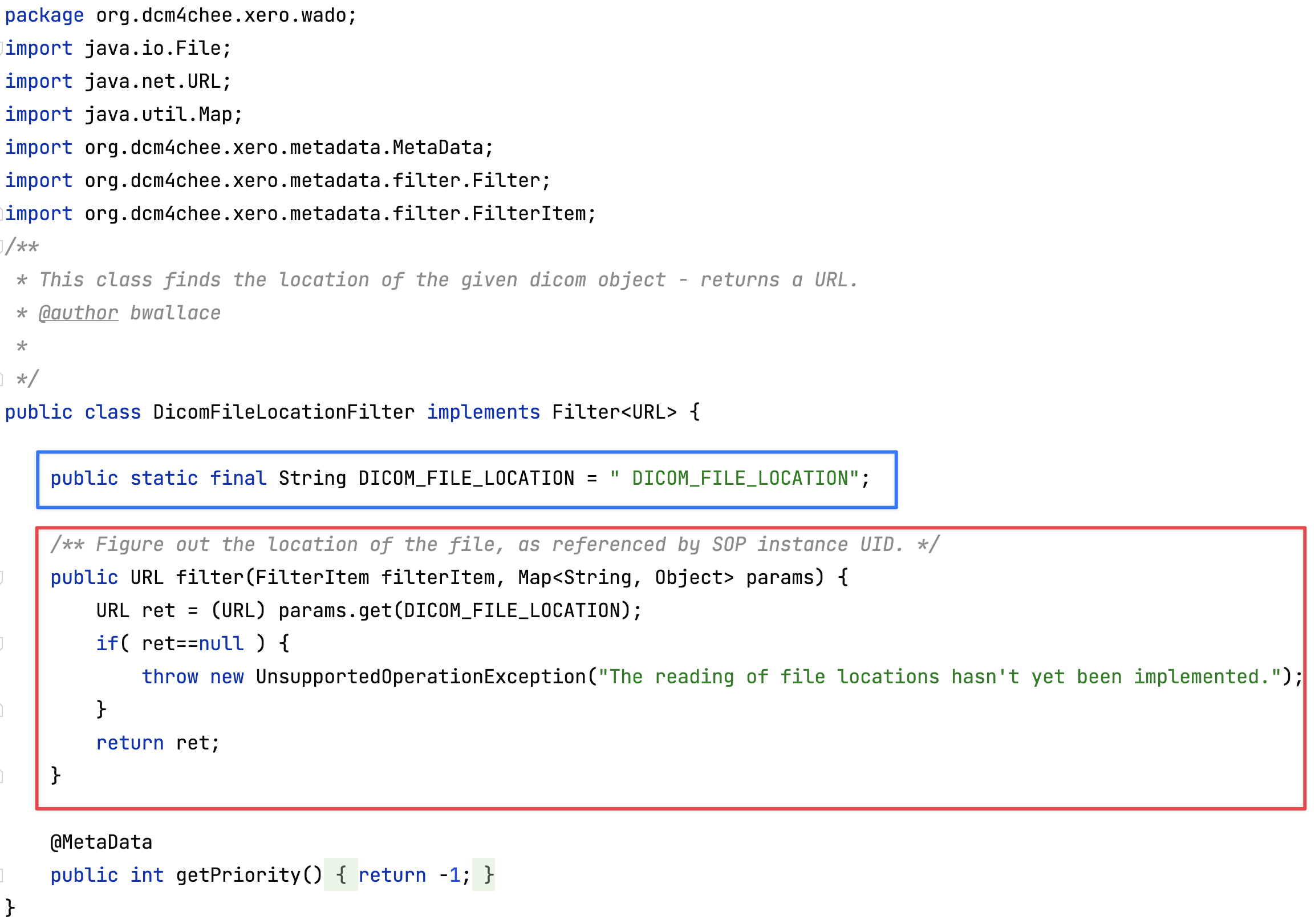}
\caption{Example of code range.}
\label{Java}
\end{figure}

Firstly, CoCC is applicable to the code-comment pair, if changes outside the code-comment pair affect the comment update, our model will not be able to detect it. In a code-comment pair, as shown in the red box in the FIGURE \ref{Java}. If the code changes in the red box affect the update of the comments in the red box, our model is applicable. But if the code changes in other places, such as the code changes in the blue box affect the update of the comments in the red box, our model is not applicable.

Secondly, the inaccuracy of the tool. Due to the large scale of the experimental data set, the quality of the data and related code features is affected by the automatic tool ChangeDistiller for change extraction.

Thirdly, the error caused by the heuristic rules used to link code and comment. When extracting the code-comment pair, we use the AST tool to link code and comment with heuristic rules, which may affect the quality of the data (especially the block-type code-comment pair). We have manually randomly checked the data to minimize the second and third threats.

Fourthly, we only choose relatively representative and common refactoring features (such as "extract method"), not all refactoring types. There are many types of code refactoring in actual development. The refactoring features we selected are designed to verify how the code refactoring affects the comment updating. It is difficult to consider all refactoring types in the actual development and extract them. We will consider more refactoring types in the future.

Lastly, when exploring the applicability of CoCC in different programming languages, due to the lack of fine-grained change extraction tools similar to ChangeDistiller in other programming languages, we collected 5000 Python samples manually. The lack of data may affect the accuracy of the experiment. However, due to the strong similarity of programming languages, outdated comments detection in other languages can be implemented similarly.

\section{CONCLUSION AND FUTURE WORK}

This paper presents a machine learning-based approach for the automatic detection of outdated comments during code changes. We selected fine-grained features of the code changes, as well as the comment features and their relation features before and after the changes.

Compared with our previous version, we have collected more open-source projects to extend the dataset. We extended the features and removed the highly correlated features, and the performance of the model was significantly improved. We also verified the applicability of CoCC in other programming languages and found outdated comments in the latest commits of open-source projects using CoCC.

In our future work, we will implement a more accurate heuristic method to link code and comment, and we will consider more refactoring features, and consider how to repair outdated comments using these features.

\appendix

\section{feature filtering\label{featurefiltering}}

Since we have collected too many features from the three dimensions of code, comments, and the relation between code and comment, there may have correlations between these features, and too many features may lead to over-fitting and affect the generalization ability of the model, so we conduct feature screening based on the previous version.

To reduce redundant features, we calculated the correlation of features and removed the features with high correlation, we use the Pearson correlation coefficient to measure the correlation between two features, as shown in formula \ref{correl}. TABLE \ref{filtering} shows the result of feature filtering.

\begin{equation}
Correl(X,Y)=\frac{\sum^{}_{} (x-\overline{x} )(y-\overline{y} )}{\sqrt{\sum^{}_{} (x-\overline{x} )^{2}\sum^{}_{} (y-\overline{y} )^{2}} } \label{correl}
\end{equation}

\begin{table}[htb]
\centering
\caption{Result of feature filtering.}
\label{filtering}
\begin{tabular}{|l|l|l|l|}
\hline
\textbf{Category} &
  \textbf{Action} &
  \textbf{Name} &
  \textbf{Reason} \\ \hline
Code features &
  Remove &
  Number of statements &
  \begin{tabular}[c]{@{}l@{}}This feature has the high correlation with the\\  proportion of code lines in the code-comment pair.\end{tabular} \\ \hline
\multirow{2}{*}{Comment features} &
  \multirow{2}{*}{Remove} &
  Length of the comment &
  \multirow{2}{*}{\begin{tabular}[c]{@{}l@{}}These two have the high correlation with the \\ proportion of code lines in the code-comment pair.\end{tabular}} \\ \cline{3-3}
 &
   &
  \begin{tabular}[c]{@{}l@{}}The ratio of comment lines to the code\\  snippet\end{tabular} &
   \\ \hline
\multirow{6}{*}{Relation features} &
  \multirow{6}{*}{Remove} &
  \begin{tabular}[c]{@{}l@{}}Similarity between old comment and \\ old code\end{tabular} &
  \multirow{2}{*}{\begin{tabular}[c]{@{}l@{}}These two features have hign correlation with the\\  distance of comment and code similarity.\end{tabular}} \\ \cline{3-3}
 &
   &
  \begin{tabular}[c]{@{}l@{}}Similarity between old comment and\\ new code\end{tabular} &
   \\ \cline{3-4} 
 &
   &
  \begin{tabular}[c]{@{}l@{}}Similarity between old comment and \\ changed statements before the change\end{tabular} &
  \multirow{2}{*}{\begin{tabular}[c]{@{}l@{}}These two features have hign correlation with the\\  distance of comment and changed statement similarity.\end{tabular}} \\ \cline{3-3}
 &
   &
  \begin{tabular}[c]{@{}l@{}}Similarity between old comment and\\  changed statements after the change\end{tabular} &
   \\ \cline{3-4} 
 &
   &
  \begin{tabular}[c]{@{}l@{}}Number of tokens old comment and \\ old code have in common\end{tabular} &
  \multirow{2}{*}{\begin{tabular}[c]{@{}l@{}}These two features have high correlation with the \\ distance of token pairs comment and code have in common.\end{tabular}} \\ \cline{3-3}
 &
   &
  \begin{tabular}[c]{@{}l@{}}Number of tokens old comment and \\ new code have in comment\end{tabular} &
   \\ \hline
\end{tabular}
\end{table}

\section{Effectiveness of word analysis features\label{wordanalysis}}

We think that the proportion of different part-of-speech in codes and comments before and after the change will affect whether the comments are outdated. Therefore, we introduce the word analysis features.

We take one of the word analysis features, “code noun token” as an example to explain the effectiveness of word analysis on detecting outdated comments. Programmers usually use noun tokens in defining variables. Feature “code noun token” can indicate the change in variables. If there are $20$ tokens in the old code, including $5$ noun tokens, then the proportion of noun tokens in the old code is $5/20=0.25$. If there are also $20$ tokens in the new code, including $8$ noun tokens, the proportion of noun tokens in the new code is $8/20=0.4$, then the value of feature “code noun token” here is $|0.4-0.25|=0.15$. We think changes in different parts of speech tokens affect outdated comments differently.

\begin{figure}[htbp]
\centering
\includegraphics[scale=0.4]{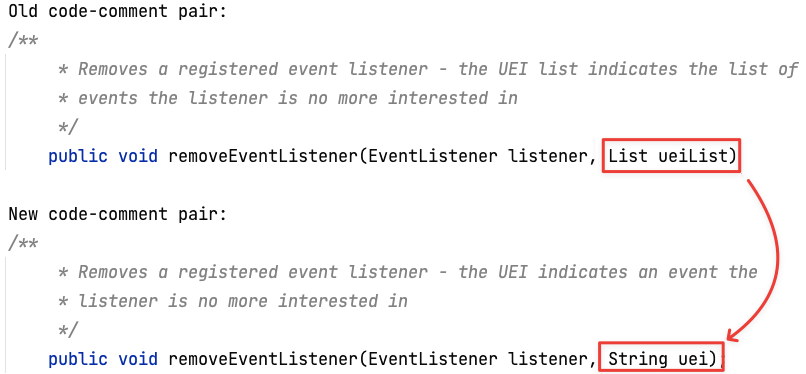}
\caption{Example of word analysis.}
\label{wordana}
\end{figure}

FIGURE \ref{wordana} shows an example, that is, the code changed, and the comment also changed. The code change occurs when the “\textbf{List ueiList}” is changed to “\textbf{String uei}”. While doing word analysis, we split the “\textbf{ueiList}” into “\textbf{uei}” and “\textbf{List}”, which are two nouns. After the change, there is only one noun “\textbf{uei}”. Therefore, the number of nouns has changed, and the comment has also changed greatly because the “\textbf{list}” is deleted, which shows the change of the noun token will affect the change of the comment.

TABLE \ref{wordanaly} is a comparison of the experimental results before and after removing word analysis features:

\begin{table}[htb]
\centering
\caption{Results with word analysis or not.}
\label{wordanaly}
\begin{tabular}{|l|l|l|l|}
\hline
\textbf{}                     & \textbf{Precision} & \textbf{Recall} & \textbf{F1} \\ \hline
All feature                   & 92.1\%             & 78.9\%          & 0.850       \\ \hline
Without word analysis  & 90.5\%             & 77.1\%          & 0.832       \\ \hline
\end{tabular}
\end{table}

After adding the word analysis feature, the model has improved, so we think word analysis is effective.

\section{window size of training word vector \label{windowsize}}

When training the word vector, we explored the training time and the effect on outdated comment detection of different window sizes. The experimental results are shown in TABLE \ref{windsize}.

\begin{table}[htb]
\centering
\caption{Training time of word vectors and detection results with different window sizes.}
\label{windsize}
\begin{tabular}{|l|l|lll|}
\hline
\multirow{2}{*}{Window size} & \multirow{2}{*}{Time(s)} & \multicolumn{3}{l|}{Outdated comment detection} \\ \cline{3-5} 
  &        & \multicolumn{1}{l|}{Precison} & \multicolumn{1}{l|}{Recall} & F1    \\ \hline
3 & 171.50 & \multicolumn{1}{l|}{91.8\%}   & \multicolumn{1}{l|}{76.6\%} & 0.835 \\ \hline
4 & 207.34 & \multicolumn{1}{l|}{91.9\%}   & \multicolumn{1}{l|}{77.3\%} & 0.839 \\ \hline
5 & 246.15 & \multicolumn{1}{l|}{92.1\%}   & \multicolumn{1}{l|}{78.9\%} & 0.850 \\ \hline
7 & 314.28 & \multicolumn{1}{l|}{92.3\%}   & \multicolumn{1}{l|}{77.7\%} & 0.843 \\ \hline
9 & 397.76 & \multicolumn{1}{l|}{92.0\%}   & \multicolumn{1}{l|}{78.2\%} & 0.845 \\ \hline
\end{tabular}
\end{table}

From the experimental results, when the window size is 5, 7, and 9, the effect on the outdated comment detection task is not much different and is better than when the window size is 3, 4. Therefore, we choose 5, which has a relatively lower training time and better performance, as the final word vector training window size.

\begin{figure}[htbp]
\centering
\includegraphics[scale=0.2]{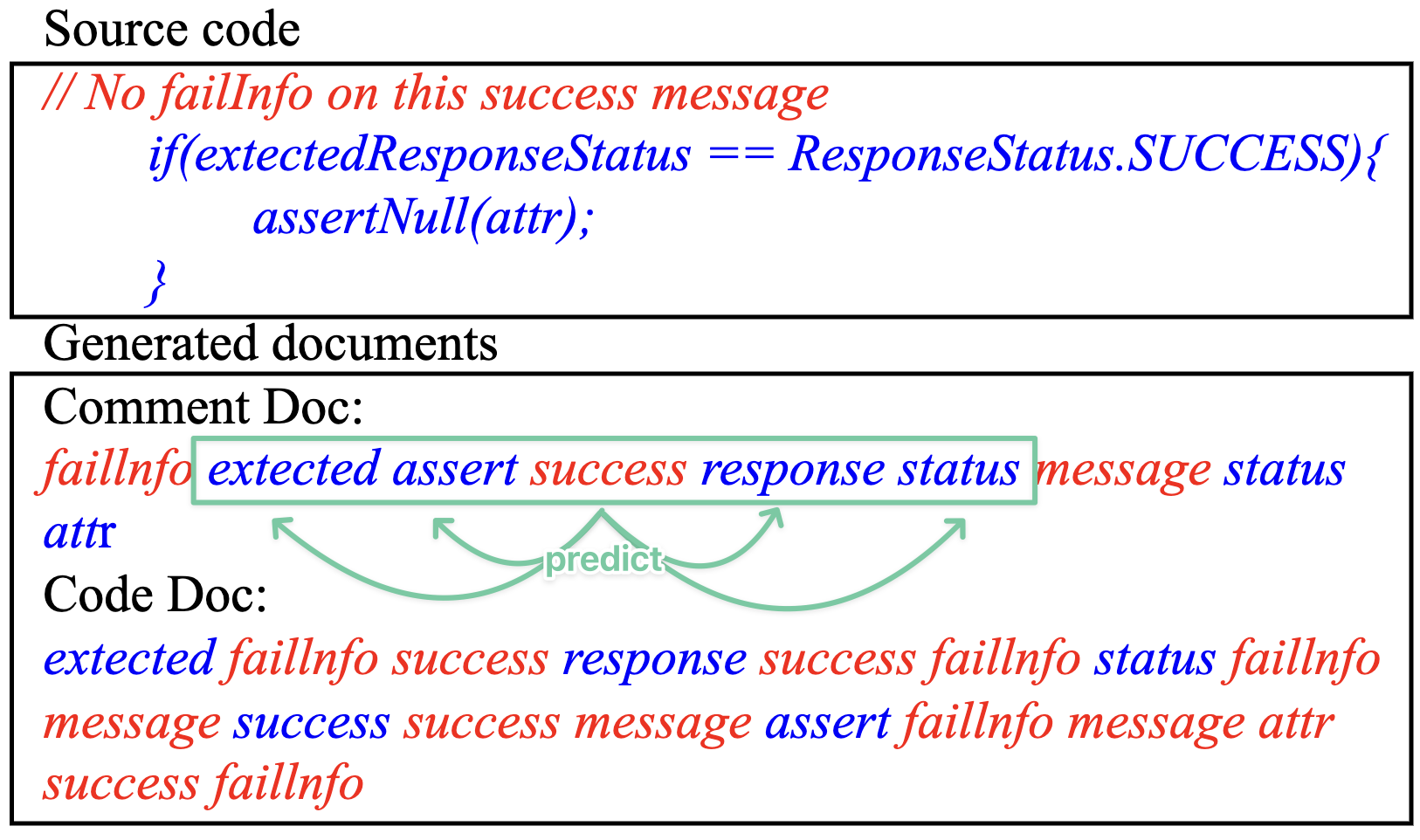}
\caption{Example of document generation.}
\label{preddoc}
\end{figure}

We think that the reason why the performance is better when the window size is 5, 7, and 9 than when the window size is 3, and 4 is related to our document (used to train word vector, detailed in Section 2.2.3) construction method. FIGURE \ref{preddoc} shows an example of document generation. We take the "Comment Doc" as an example ("Code Doc" is the same), in the green box, the central word "success" comes from the comment, and the four words around it come from the code. When training the word vector, the sliding window size is fixed, so intuitively, when the window size is greater than or equal to 5, the token from the comment and code can be covered. Therefore, the performance is better when the window size is 5, 7, and 9 than when the window size is 3, and 4.

\section{Feature about similarity between code and comment \label{smlt}}

When we calculate the relation features between code and comment, we first use formula \ref{eq3} to formula \ref{eqc} to calculate the required similarity between code and comment. The similarity feature calculated by this method is a decimal value between 0 and 1. In order to explore the impact of using the original feature values (decimal values) and artificially defined high and low similarity thresholds (for example, when the threshold is 0.1, if the similarity exceeds 0.1, it will be considered as high similarity), on the outdated comment detection task, we conducted the following experiments: iterating different thresholds in steps of 0.1, using only relation features (because similarity features are all in relation features) to detect outdated comments. TABLE \ref{ster} shows the experimental results.

\begin{table}[htb]
\centering
\caption{Similarity thresholds experimental results.}
\label{ster}
\begin{tabular}{|l|l|l|l|}
\hline
\textbf{Threshold} & \textbf{Precision} & \textbf{Recall} & \textbf{F1}    \\ \hline
0.1                & 77.5\%             & 56.3\%          & 0.652          \\ \hline
0.2                & 79.3\%             & 40.2\%          & 0.534          \\ \hline
0.3                & 72.7\%             & 35.3\%          & 0.475          \\ \hline
0.4                & 72.5\%             & 28.2\%          & 0.405          \\ \hline
0.5                & 72.5\%             & 26.1\%          & 0.384          \\ \hline
0.6                & 64.3\%             & 32.6\%          & 0.433          \\ \hline
0.7                & 66.0\%             & 31.9\%          & 0.430          \\ \hline
0.8                & 66.8\%             & 31.5\%          & 0.428          \\ \hline
0.9                & 66.8\%             & 31.2\%          & 0.425          \\ \hline
\textbf{None}      & \textbf{84.0\%}    & \textbf{66.6\%} & \textbf{0.743} \\ \hline
\end{tabular}
\end{table}

From the experimental results, the outdated comment detection task without setting the threshold of high and low similarity has the best effect. Therefore, the threshold trained from a large number of samples through random forest or other algorithms is better than that set manually, and the performance of the model is better as well.

\section{Fine-tuning of the hyperparameters\label{hyperparameters}}

To achieve the best performance of different models, we used 10-fold-cross-validation and grid search to select the best hyperparameters in the experiment, the grid search algorithm is a parameter fine-tuning method, which optimizes the model performance by traversing the given combination of parameters. We split all data (details in TABLE \ref{dataset}) into the training set and test set at a ratio of 7:3, and then perform hyperparameter fine-tuning on the training set. Specifically, we divide the training set into 10 parts on average, taking the $K_{th}$ part as the cross-validation set each time, and the remaining $9$ parts as the training set.

TABLE \ref{hal} shows the hyperparameter list and the best choice.

\begin{table}[htb]
\centering
\caption{Hyperparameter fine-tuning result.}
\label{hal}
\begin{tabular}{|l|l|l|l|}
\hline
\textbf{Models}                      & \textbf{Hyperparameters} & \textbf{Alternative values}              & \textbf{Best one} \\ \hline
Naive bayes (GaussianNB)             & var\_smmothing           & 1e-9, 1e-8, 1e-7, 1e-6, 1e-5 & 1e-5              \\ \hline
\multirow{3}{*}{SVM (SVC)}           & C                        & 0.1, 1, 10, 100              & 0.1               \\ \cline{2-4} 
                                     & gamma                    & 0.1, 1, 10, 100              & 0.1               \\ \cline{2-4} 
                                     & Kernel                   & ‘linear’, ‘rbf’              & ‘rbf’             \\ \hline
\multirow{3}{*}{Logistic regression} & penalty                  & ‘l1’, ‘l2’                   & ‘l1’              \\ \cline{2-4} 
                                     & C                        & 0.001, 0.01, 0.1, 1, 10, 100 & 100               \\ \cline{2-4} 
                                     & solver                   & ‘liblinear’, ‘saga’          & ‘liblinear’       \\ \hline
\multirow{4}{*}{Decision tree}       & criterion                & ‘gini’, ‘entropy’            & ‘gini’            \\ \cline{2-4} 
                                     & max\_depth               & None, 5, 10, 20              & 20                \\ \cline{2-4} 
                                     & min\_samples\_split      & 2, 5, 10                     & 5                 \\ \cline{2-4} 
                                     & min\_samples\_leaf       & 1, 2, 4                      & 1                 \\ \hline
\multirow{6}{*}{Random forest}       & n\_estimators            & 50, 100, 200, 300            & 200               \\ \cline{2-4} 
                                     & criterion                & ‘gini’, ‘entropy’            & ‘gini’            \\ \cline{2-4} 
                                     & max\_depth               & None, 5, 10, 20              & None              \\ \cline{2-4} 
                                     & min\_samples\_split      & 2, 5, 10                     & 2                 \\ \cline{2-4} 
                                     & min\_samples\_leaf       & 1, 2, 4                      & 1                 \\ \cline{2-4} 
                                     & max\_features            & ‘sqrt’, ‘log2’, None         & ‘sqrt’            \\ \hline
\multirow{5}{*}{XGBoost}             & learning\_rate           & 0.01, 0.1, 1                 & 1              \\ \cline{2-4} 
                                     & max\_depth               & 3, 5, 7                      & 7                 \\ \cline{2-4} 
                                     & n\_estimators            & 50, 100, 200                 & 200                \\ \cline{2-4} 
                                     & subsample                & 0.5, 0.7, 1.0                & 1.0               \\ \cline{2-4} 
                                     & colsample\_bytree        & 0.5, 0.7, 1.0                & 1.0               \\ \hline
\end{tabular}
\end{table}

\bibliography{wileyNJD-AMA}%

\end{document}